\documentclass[fleqn,usenatbib]{mnras}

\usepackage{newtxtext,newtxmath}
\usepackage{lscape}
\usepackage{array}
\usepackage{graphics}
\usepackage{tikz}
\usepackage{xcolor}
\usepackage{orcidlink} 
\usepackage{upgreek}
\usepackage{textcomp}
\usepackage{subfigure}
\usepackage{subcaption}

\usepackage[T1]{fontenc}

\newcommand{\hmshh}{\ensuremath{^\mathrm{h}}}
\newcommand{\hmsmm}{\ensuremath{^\mathrm{m}}}
\newcommand{\hmsss}{\ensuremath{^\mathrm{s}}}
\newcommand{\degree}{\ensuremath{^{\circ}}}
\newcommand{\arcm}{\ensuremath{^{\prime}}}
\newcommand{\arcs}{\ensuremath{^{\prime\prime}}}

\newcommand{\frb}{FRB20250613A}
\def\code#1{\texttt{#1}}

\DeclareRobustCommand{\VAN}[3]{#2}
\let\VANthebibliography\thebibliography
\def\thebibliography{\DeclareRobustCommand{\VAN}[3]{##3}\VANthebibliography}

\usepackage{graphicx}	
\usepackage{amsmath}	

\title[FRB20250613A]{FRB20250613A: a remarkable repeating FRB with apparent millisecond-timescale scattering variations}

\author[T. Dial et al.]{
T. Dial,$^{1}$\thanks{E-mail: tdial@swin.edu.au}
A. T. Deller, $^{1}$
Alexa~C.~Gordon \orcidlink{0000-0002-5025-4645}, $^{2}$
P.~A.~Uttarkar, $^{1}$
R. M. Shannon \orcidlink{0000-0002-7285-6348}, $^{1}$
Ziteng Wang \orcidlink{0000-0002-2066-9823}, $^{3}$
\newauthor
M. Caleb, $^{4,5}$
Wen-fai Fong, $^{2}$
Marcin Glowacki, $^{6,7}$
Kelly Gourdji, $^{8}$
Joscha N.\ Jahns-Schindler \orcidlink{0000-0003-4193-6158}, $^{1}$
\\
$^{1}$Center for Astrophysics and Supercomputing, Swinburne University of Technology, P.O. Box 218, Hawthorn, Vic 3122 Australia
\\
$^{2}$Center for Interdisciplinary Exploration and Research in Astrophysics (CIERA) and Department of Physics and Astronomy, \\ Northwestern University, Evanston, IL 60208, USA
\\
$^{3}$International Centre for Radio Astronomy Research, Curtin University, Bentley, WA 6102, Australia
\\
$^{4}$Sydney Institute for Astronomy, School of Physics, The University of Sydney, NSW 2006, Australia
\\
$^{5}$ARC Centre of Excellence for Gravitational Wave Discovery (OzGrav), Hawthorn, 3122, Victoria, Australia
\\
$^{6}$Institute for Astronomy, University of Edinburgh, Royal Observatory, Edinburgh, EH9 3HJ, United Kingdom
\\
$^{7}$Inter-University Institute for Data Intensive Astronomy, Department of Astronomy, University of Cape Town, Cape Town, South Africa
\\
$^{8}$Australia Telescope National Facility, CSIRO, Space and Astronomy, PO Box 76, Epping, NSW 1710, Australia
}

\date{Accepted XXX. Received YYY; in original form ZZZ}

\pubyear{\the\year{}}

\begin{document}
\label{firstpage}
\pagerange{\pageref{firstpage}--\pageref{lastpage}}
\maketitle

\begin{abstract}
Fast Radio Bursts (FRBs) are bright millisecond-duration, extragalactic radio bursts whose impulsive emission can be used to probe diffuse ionised gas. With appropriate modelling, this can be translated into constraints on large scale structure and cosmological models. However, an important and as-yet poorly constrained component of these models is the burst progenitor and local environment. In this regard, repeating FRBs provide strong insight as their successive bursts can reveal changes in the line of sight to the burst progenitor on short timescales, with several sources showing evidence for a progenitors embedded in dense, turbulent, and highly magnetised environments. FRB20250613A is a repeating FRB discovered by the Australian SKA Pathfinder and localised to a low-metallicity dwarf galaxy at a redshift of $z = 0.0987 \pm 0.0001$. FRB 20250613A exhibits a plethora of exotic features  that likely overlay the imprint of the circum-burst environment on some intrinsic features of the source. Here we perform a comprehensive analysis of bursts detected by ASKAP, MeerKAT, and the Murriyang Parkes radio telescopes. Bursts during the MeerKAT epoch show a large apparent variance in scattering on timescales of minutes to hours. Polarimetric analysis of the full sample shows spectral depolarisation with variability on timescales of days and changes in rotation measure of $\sim$ 300 rad m$^{-2}$ over days to months. This suggests a highly turbulent magneto-ionised environment. We find significant preference for separations of $\sim$6.8$\pm$0.8 ms in multi-component bursts that we suggest is likely intrinsic to the burst emission mechanism. Finally, we find that a subset of bursts exhibit variations in these propagation effects on burst components separated by just milliseconds, that are difficult to explain by changing sightlines, but plausibly due to non-linear plasma effects in the circum-burst environment caused by the high field strength of the FRB emission. These properties, which demand a nearby turbulent screen of material, are all consistent with the FRB progenitor being embedded in the dense stellar wind of a Be star binary companion, objects which are relatively plentiful in low-mass and low-metallicity galaxies like the FRB\,20250613A host.

\end{abstract}




\section{Introduction}
Fast Radio Bursts (FRBs) are extremely luminous, (sub-)millisecond-duration radio bursts from extragalactic sources \citep{lorimer2007bright}. They are especially useful for their high susceptibility to propagation effects such as dispersion due to free-electrons, which can inform us about the matter content and large-scale structure in the universe \citep{macquart2020census}. However, the types of objects and environments that can support FRBs remain largely unknown; this uncertainty hinders our ability to model their populations, thereby limiting their potential as cosmological probes.

Recent advancements by facilities such as the Australian SKA Pathfinder \cite[ASKAP, ][]{hotan2021australian, wang2025craft,shannon2024commensal}, Canadian HI Intensity Mapping Experiment \cite[CHIME, ][]{amiri2022overview, lanman2024chime, amiri2025chime}, and Deep Synoptic Array \cite[DSA][]{ravi2023deep} have established a large catalogue of FRBs, consisting of $\sim$4000 unique sources \citep{abbott2026second}. Many exhibit complex time and frequency morphologies as well as high linear polarisation fractions that are reminiscent of pulsar and magnetar emission \citep{pandhi2024polarization}. Magnetars are thought to produce at least a subset of FRBs as FRB-like emission was detected from the Galactic magnetar SGR\,1935+2154 \citep{chime2020bright, bochenek2020fast}. The extreme luminosities of some cosmological FRBs suggest they may be powered by highly energetic magnetar events, although the precise mechanism remains uncertain \citep{wu2025universal}. The vast majority of bursts are observed as isolated one-off events, which, ignoring selection effects, suggest that some fraction could arise from cataclysmic events such as compact mergers \citep[e.g.][]{wang2016fast}. However, $\sim$100 FRB sources have been seen to emit repeat bursts \citep{abbott2026second}, some of which are very active, with thousands of bursts seen to date \citep{zhang2025magnetar}.

The ability to conduct targeted follow-up observations of repeating FRBs across different frequency bands and observing configurations enables detailed characterisation of the source and local environment. These have been shown to vary drastically between sources, and are often dynamic and time varying within a given source. The Rotation Measure (RM), for instance, has been shown to vary greatly from source to source, but also between bursts of a given source. The longest-studied source, FRB~20121102A, reported the largest magnitude RM \cite[$\sim$10$^{5}$ rad m$^{-2}$,][]{michilli2018extreme} and has steadily declined by $\sim$15 per cent per year, which could support a compact object embedded in a supernovae remnant \citep{hilmarsson2021rotation}. The extreme 10$^{4}$ rad m$^{-2}$ variations observed for FRB~20190520B has been attributed to compact binary interactions \citep{anna2023magnetic}. Additionally, a potential coronal mass ejection (CME) may have crossed the line of sight (LOS) towards FRB~20220629A, resulting in a large temporary fifty-fold increase in RM \citep{li2026sudden}. However, it is still unclear the extent to which the environments around repeating FRBs (especially highly active ones) can be generalised to the broader FRB population. 

Spectral depolarisation is another common feature in repeating FRBs that supports a dense turbulent ionised medium close to the burst progenitor. Multipath propagation through this medium results in depolarisation that increases towards longer wavelengths as the number and variation of multiple paths taken by the pulse grows. Using a sample of repeating FRBs, \cite{feng2022frequency} derived empirical power-law models describing the relation between the strength of the spectral depolarisation, characterised by the RM scatter in the plasma (denoted by $\sigma_{\mathrm{RM}}$) and both the scattering timescale $\tau$ and $|\mathrm{RM}|$. Given that in several repeating FRB sources the RM has been seen to evolve over time, temporal variations in $\sigma_{\mathrm{RM}}$ for individual sources are also expected. This was shown for FRB20201124A \citep{lu2023temporal} where the evolution of depolarisation and magnetic field traced the RM, suggesting these variations are caused by the same magnetoionised environment. Spectral depolarisation is absent and weaker in apparently one-off FRBs \cite[][]{2024MNRAS.527.4285U,2026MNRAS.545f1997U}, although current analyses are sensitive to only a limited range of $\sigma_{\mathrm{RM}}$ values.

Additionally, instances of large variations in the scattering timescale have been observed for a number of sources. In \cite{ocker2023scattering} and \cite{panda2025low}, significant changes in scattering were observed over minute-hour timescales for FRB20190520B and FRB20240114A, respectively. Additionally, scattering time variations on timescales of days have been observed for FRB20180916B \citep{sand2023chime, gopinath2024propagation}. \cite{ocker2023scattering} suggests this could be caused by turbulent and patchy environments on small spatial scales close to the FRB source. 

These propagation effects can mask aspects of the underlying FRB emission mechanism, which offers complementary insight into the FRB progenitor. However, one aspect which is not greatly affected is the wait time distribution. In many repeating FRBs, a bimodal wait time is seen \citep{zhang2018fast, zhang2022fast, zhang2023fast, zhang2025magnetar} with a sub-second peak that may represent multiple components of an individual burst and a second peak at $\gg1$ second. Although it becomes difficult to differentiate multi-component bursts from genuinely individual bursts, the sub-second waiting times implicate the source of the emission as a compact object such as a neutron star (NS). However, the broadness of the distribution requires masking of any strict periodicity through processes such as stochastic beaming or different emission heights \citep{li2021bimodal}.

In this paper, we report on FRB 20250613A, a repeating FRB discovered and localised by ASKAP and observed with both the MeerKAT and Murriyang/Parkes radio telescopes. We present a compilation of detected multi-component bursts in Fig. \ref{fig:dynspec_multicomp_mosaic}. The bursts exhibits a variety of temporal, spectral and polarimetric properties. Many of these properties have been observed in other repeating FRBs, but some are entirely novel, and the combination of which has not been seen in a single source. In Section \ref{sec: METHODS part1} we describe the detection and follow-up observations of FRB~20250613A, including its  localisation to a low mass dwarf galaxy,  and an analysis of its properties and host environment. Section \ref{sec:Methods part2} discusses the burst-to-burst variations in RM, $\sigma_{\mathrm{RM}}$ and $\tau$. In Section \ref{sec:DISCUSSION} we consider the temporal variations of $\tau$, RM and $\sigma_{\mathrm{RM}}$ in the context of a Be star / NS model with a strong stellar wind. We show that for the parameters that fit the longer-timescale evolution of $\tau$, RM, and $\sigma_{\mathrm{RM}}$, the bursts can induce non-linear propagation effects as they pass through the stellar wind, which would be qualitatively consistent with the changes in pulse broadening that occur over millisecond timescales that are difficult to explain otherwise. We highlight a strongly preferred component separation that has not been seen in other repeating FRBs, which combined with the bursts polarisation signature supports a neutron star origin. Finally, we present our conclusion in Section \ref{sec:CONCLUSION}. All modelling (unless specified otherwise) was done with the Dynesty (Dynamic nested sampling) Sampler \citep{speagle2020dynesty} with the use of the \textsc{Bilby} \citep{ashton2019bilby} package, a Bayesian inference python library. All uncertainties are reported to 68 per cent confidence.

\section{Observations and burst analysis}
\label{sec: METHODS part1}

\subsection{Radio observations and analysis}

\subsubsection{ASKAP}
\label{sec: ASKAP_burst_detection}

On 13 June 2025, at UTC 22:21:51 (MJD = 60839.9318351840), the incoherent summation (ICS) FRB detection system \citep{shannon2024commensal} on ASKAP \citep{hotan2021australian} reported the discovery of \frb\ as part of the Commensal Real-Time ASKAP Fast-Transients collaboration \citep[CRAFT;][]{macquart2010commensal}. The FRB was detected in the ASKAP low band, with 336 MHz of observing bandwidth centred at 919.5 MHz, a reported dispersion measure (DM) of 174.87 pc\,cm$^{-3}$ and a boxcar width of 3 samples corresponding to 3.546 ms. A download of the full 12.4\,s of 1-bit voltage data was triggered by the real-time detection system and transferred to the Ngarrgu Tinderbeek (NT) supercomputer at Swinburne University of Technology. Calibration observations of the polarisation calibrator PSR J0835$-$4510 (Vela) and bandpass calibrator, the radio galaxy PKS B0407$-$658, were taken 9.8 and 10 hours respectively after the initial FRB detection; similarly with 12.4\,s of 1-bit voltage data. The raw voltages of \frb\ and the calibrator sources were then processed using the CRAFT Effortless Localisation and Enhanced Burst Inspection (CELEBI) pipeline \citep{scott2023celebi, glowacki2026} to produce an accurate sub-arcsecond position of 04\hmshh36\hmsmm33.04\hmsss -44\degree31\arcm58.57\arcs with uncertainty 0.55\arcs $\times$ 0.44\arcs with a position angle of 74.37 deg and a structure-maximised DM \citep[SMDM;][]{sutinjo2023calculation} of 174.60 $\pm$ 0.04 pc cm$^{-3}$. This was used to produce coherently de-dispersed beamformed high time resolution (HTR) dynamic spectra with full calibrated polarimetry as detailed in \cite{dial2025frb}.

An additional burst from FRB20250613A was detected on 26 October 2025 at UTC 16:55:03 using the ICS FRB system. This burst was similarly detected in the "low" band with a central frequency of 863.5 MHz. The same polarisation and bandpass calibrators were observed 2.0 and 2.2 hours respectively after the burst detection. The raw voltages were transferred to the NT supercomputer and processed using the same procedure as for the first burst.

The first ASKAP burst A1 (see Fig.~\ref{fig:dynspec_multicomp_mosaic}) shows two distinct components. The leading component is broader in time and occupies a higher frequency region compared to the sharper and brighter trailing component. Since the noise floor separates the two components (by $\sim$2 ms) we treat them as separate bursts in the following analysis. The two components were coherently de-dispersed to a SMDM of 174.88 $\pm$ 0.66 pc cm$^{-3}$ and 174.61 $\pm$ 0.05 pc cm$^{-3}$ respectively using the \textsc{SHRINE}\footnote{SHRINE github: https://github.com/marcinglowacki/SHRINE} codebase \citep{sutinjo2023calculation}.

\begin{figure*}
    \centering
    \includegraphics[width=0.8\textwidth]{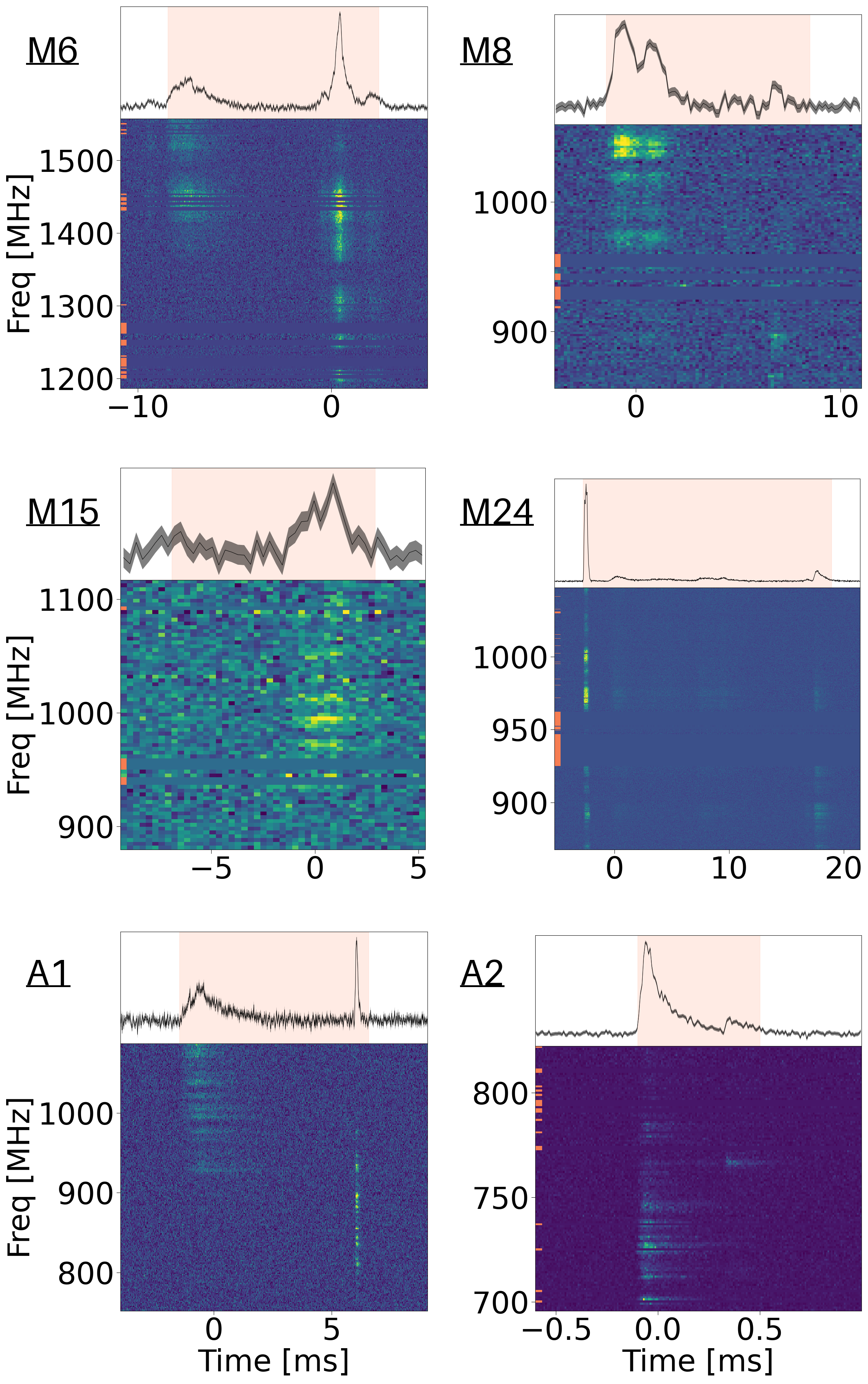}
    \caption{Multi-component bursts from FRB20250613. The letter in front of the burst number represents the telescope which detected the burst i.e. `A' for ASKAP, `M' for MeerKAT and `P' for Parkes (Murriyang). Top panel: Frequency-scrunched time series. The shaded grey band shows the off-pulse sample noise. The red region shows the on-pulse width. The burst reference time is set to the centroid of the burst. Bottom panel: Dynamic spectrum. Flagged channels are denoted by the orange markings.}
    \label{fig:dynspec_multicomp_mosaic}
\end{figure*}

\begin{figure*}
    \ContinuedFloat
    \centering
    \includegraphics[width=0.8\textwidth]{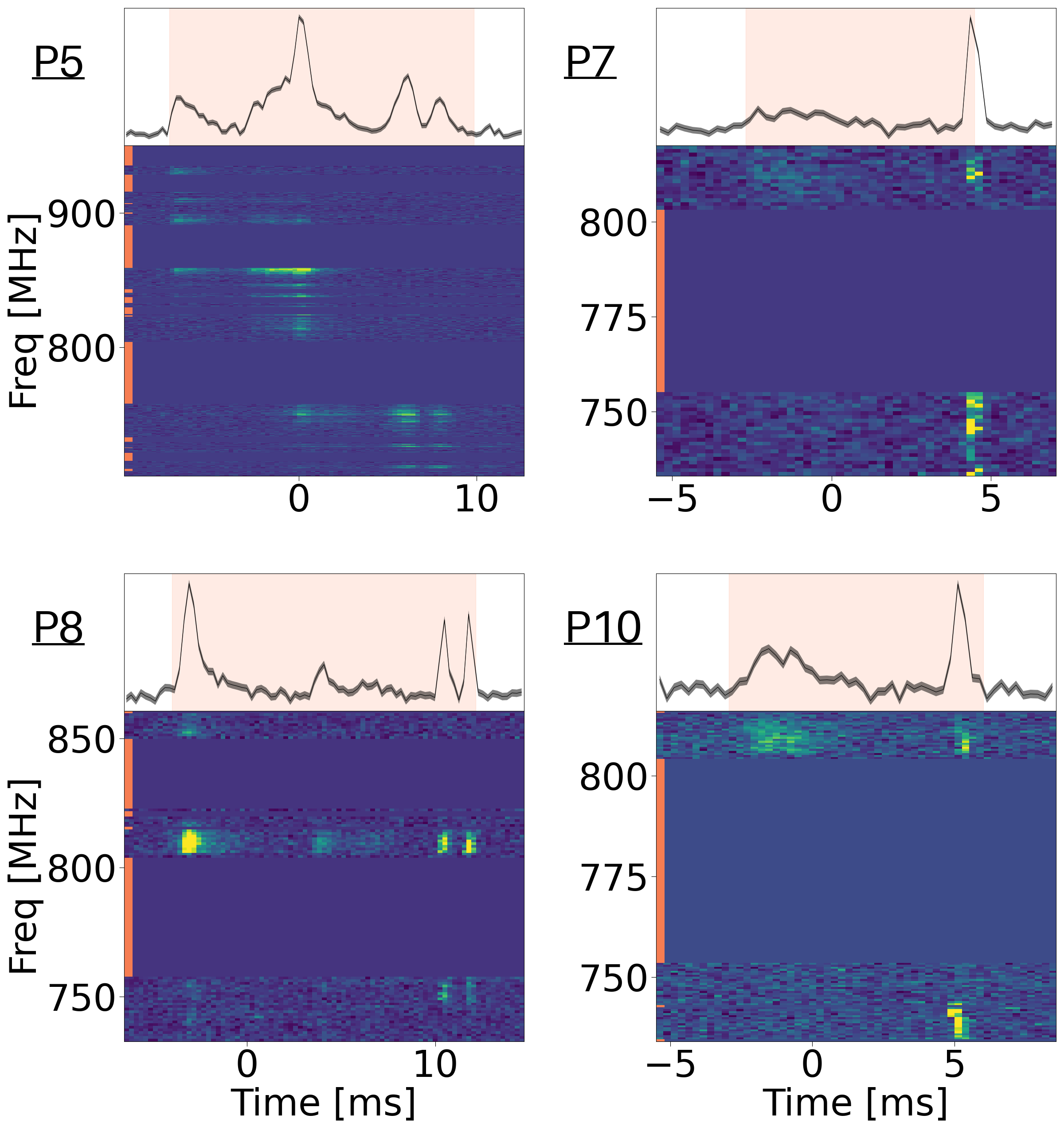}
    \caption{(cont.)}
\end{figure*}

\subsubsection{MeerKAT}
\label{sec: MEERKAT_desc}

Follow up MeerKAT observations of FRB~20250613A were carried out on the 23-24 June 2025 at the CELEBI-derived sky position as part of the DDT proposal DDT-20250621-AW-01 using the Transient User Supplied Equipment (TUSE) and Pulsar Timing User Supplied Equipment \cite[PTUSE,][]{bailes2020meerkat} . A total of 8 hours of observations were taken with the L-band (856-1712 MHz) receivers, 2 hours on 23 June with 58 of the 64 13.5-m dishes and 4 hours on 24 June with 57 dishes, formed with 768 coherently added tied-array beams. The  PTUSE observations were coherently dedispersed at a nominal DM of 174.6 pc cm$^{-3}$, recorded  in \textsc{PSRFITS} \citep{hotan2004psrchive} format, and then transferred to the NT supercomputer for offline burst searching. To search for potential bursts, we used the GPU-based single-pulse search software \code{heimdall}. For classification we used the \textsc{FETCH} classifier with model A \citep{agarwal2020fetch}. The candidates were filtered within a DM range of 170-182 pc\,cm$^{-3}$ and a threshold S/N of greater than 7.5 before being inspected manually. In total, 24 bursts were detected, two on June 23 and 22 on June 24. The full 8-s filterbank file containing the bursts were converted to Stokes dynamic spectra using \textsc{sigproc}\footnote{https://sigproc.sourceforge.net/}. For the 23 June observations, observations were recorded with 4096 channels across the band, at a temporal resolution of 43.07 \textmu s for observations; for the 24 June observations we recorded with  1024 channels at 38.28 \textmu s resolution. Voltage triggers \citep{rajwade2024study} were available for most of the detected bursts. For the work presented here, only the voltages for Burst M24 were transferred to the NT supercomputer for offline analysis. This was the brightest MeerKAT burst, meaning that very high time resolution analysis was possible, and moreover it displayed significant artefacts in the search mode data that we could mitigate through use of the voltage data. The voltages were manually beamformed using the \textsc{TBeamformer}\footnote{https://gitlab.com/kmrajwade/tbeamformer} codebase, after which Stokes filterbanks were created with 4096 frequency channels at 5 \textmu s time resolution.

\subsubsection{Murriyang (Parkes)}
\label{sec: PARKES_desc}
 After the discovery of the source with ASKAP we triggered observations from Murriyang (project ID P1343). The source was observed with the ultrawideband low (UWL) receiver \cite[][]{2020PASA...37...12H}, which observes in  a band spanning 704-4032 MHz, at 20 epochs, with a total integration time of $\sim$32 hours, with the last epoch on November 8, 2025. While the initial four epochs of observation from Murriyang after the discovery of the source did not result in any detection, a long-term monitoring campaign from Murriyang (project ID P1344) showed renewed activity starting on the  18 August 2025 epoch. The properties of all 36 bursts are given in Table \ref{tab:burst_tab1}.

The source was observed in pulsar search mode, with a temporal and spectral resolutions of 256 $\upmu$s and 0.5 MHz respectively. We use a \code{heimdall}-based subbanded search method (see \citealt{Kumar:2021} for a detailed description) to search for FRB emission. We use 13 different subbands of varying spectral widths from 3328 MHz to 64 MHz to search for dispersed candidates. 
We searched the data over a DM range of 100 and 900 ${\rm pc\,cm^{-3}}$. The candidates identified by the \code{heimdall}-based search are classified using \code{fetch} to identify credible candidates. We also perform a manual classification of candidates between a DM of 160 $\rm {pc\,cm^{-3}}$ and 180 $\rm {pc\,cm^{-3}}$ in addition to the \code{fetch}-classified candidates, to ensure no bona-fide candidates are missed.

\subsubsection{Burst analysis: identification of on-burst region}
\label{sec: findfrb}

We improved the methods used in \cite{scott2025high} to identify the on-pulse region of each burst in both time and frequency, as well as filter out channels with significant time-variable RFI. The routine is as follows. 

\textbf{(1) Conservative RFI flagging}: We perform flagging using a statistical approach where the standard deviation of each channel, $\mathrm{\sigma_{t}}(f)$, is compared to the median value of the full bandwidth. Channels are filtered out if they exceed a specific threshold $M$:

\begin{equation}
\label{eq:filter_conditions}
\begin{split}
    & I(f) = 
    \begin{cases}
        0 & \hspace{0.5cm}
        \begin{split}
        & \hspace{0.8cm}\Big|\sigma_{t}(f) - \mathrm{med}\big(\sigma_{t}(f)\big)\Big|
        > \\ &M  \cdot \mathrm{med}\Big(\Big|\sigma_{t}(f) - \mathrm{med}\big(\sigma_{t}(f)\big)\Big|\Big)
        \end{split}
        \\
        I(f) & \hspace{2.0cm}\mathrm{otherwise},
    \end{cases}
\end{split}
\end{equation}

where $\sigma_{t}(f) = \sqrt{\sum_{t}\mathrm{I}(f,t)^{2}}$. For each burst, the Stokes \textit{I} dynamic spectrum is down-sampled to a resolution of 1 ms before performing RFI flagging. This algorithm is performed over 5 iterations. After each iteration the filtering conditions in equation \ref{eq:filter_conditions} are re-calculated to remove bad RFI channels; This ensures both minor and major affected channels are removed. The data are then band-averaged to locate the peak of the burst. 

\textbf{(2) Temporal Bounding}: A window with a width of $\mathrm{w_i}$ (defaulting to 30.0 ms) centred on the peak of the burst is taken from the native resolution dynamic spectrum. We use a width minimisation algorithm to obtain the temporal bounds of the burst \citep{scott2025high}. The data are band-averaged and convolved with a window of sample length N$\mathrm{_{t}}$ = 2 starting from the first time sample. The sample position p$\mathrm{_{t}}$ and the window length N$\mathrm{_{t}}$ are recorded when the window fluence exceeds 95 per cent of the total integrated fluence:

\begin{equation}
\sum^{\mathrm{p_{t}+N_{t}}}_{\mathrm{p_{t}}}I(t) > \mathrm{y_{frac}}\sum I(t),
\end{equation}

where y$_{\mathrm{frac}}$ = 0.95 is the fractional fluence threshold. If this condition is not met, N$\mathrm{_{t}}$ is incremented until an appropriate pair, (p$\mathrm{_{t}}$, N$\mathrm{_{t}}$), is found which is then used to define the on-pulse region in time; If multiple values of p are found, the median value is used.

\textbf{(3) Matched Filtering}: A new window based on (p$\mathrm{_{t}}$, N$\mathrm{_{t}}$) is extracted from the dynamic spectrum and band averaged, and then smoothed using a Gaussian kernel to obtain weights $\mathrm{W}_t$. These are applied as a matched filter when time-averaging to obtain the frequency spectrum: $I(f) = \mathrm{1/N} \cdot\sum_{t}^{\mathrm{N}}(\mathrm{W_{t}}\cdot I(f, t))$. 

\textbf{(4) Spectral Bounding}: The width-minimisation algorithm is repeated on $I(f)$ to find the frequency bounds (p$\mathrm{_{f}}$, N$\mathrm{_{f}}$).

\textbf{(5) RFI Subtraction}: The rough on-pulse region defined by (p$\mathrm{_{t}}$, N$\mathrm{_{t}}$) and (p$\mathrm{_{f}}$, N$\mathrm{_{f}}$) is used to subtract the baseline using the mean of the off-pulse region. Two background windows of width $\mathrm{w_{rms}/2}$ (default 5.0 ms) are taken from the native resolution dynamic spectrum, flanking the on-pulse region. These are time-averaged per channel and subtracted from the full dynamic spectrum. 

\textbf{(6) Iterative Refinement}: A new window of width $\mathrm{w_i}$, centred on the burst peak, is taken from the RFI-subtracted dynamic spectrum.  Channels outside the on-pulse region are removed, and steps 1-4 are repeated to optimise the on-pulse parameters. The final output is a cropped, RFI-subtracted, Stokes \textit{I} dynamic spectrum.

The width minimisation algorithm is highly sensitive to RFI and baseline noise, necessitating the conservative RFI flagging and matched filtering employed here. For most bursts in the sample, $\mathrm{y_{frac} =}$ 0.95 fully captured the on-pulse region, provided the integrated signal-to-noise (S/N) was $\gtrsim$20.0. For fainter bursts, a lower threshold ($\mathrm{y_{frac}}=$ 0.80-0.90) was used. In cases of insufficient S/N where the algorithm failed, cropping and RFI removal were performed manually. A threshold of $\mathrm{y_{frac}} >$0.95 was used for extremely bright bursts. The final estimated temporal widths and $\mathrm{y_{frac}}$ values recorded for each burst are shown in Table~\ref{tab:burst_tab1} in the appendix.

\subsubsection{Fitting burst morphology and scattering}
\label{sec: temp_analysis}

The Stokes \textit{I} dynamic spectrum for each burst was band-averaged and down-sampled to improve the S/N. A similar method to that detailed in \cite{dial2025frb} was used to fit the morphology. Briefly, we assume that the intrinsic structure of the burst can be modelled as a sum of Gaussian pulses. As the burst propagates through a turbulent cold plasma, it undergoes multi-path propagation, resulting in the intrinsic burst being convolved with a one-sided exponential scattering tail characterised by a scattering timescale $\tau$:

\begin{equation}
    \label{eq:scattering}  
    I(t) = \sum_{i = 1}^{N}\bigg[A_{i}e^{-(t-\mu_{i})^{2}/2\sigma_{i}^{2}}\bigg] * e^{-t/\tau},
\end{equation}

where $A_{i}$, $\mu_{i}$ and $\sigma_{i}$ are the amplitude, position and width of each Gaussian pulse, respectively, and `$*$' denotes the convolution operator. This assumes a Gaussian scattering disk and we explore the limitation of this in Section \ref{sec:scatter_var}. Since the number of Gaussian pulses required to represent each burst is unknown, we begin with a single pulse using wide priors. Additional Gaussian components are iteratively added and evaluated against previous models through review of two statistical metrics: the Bayesian Information Criterion (BIC) and Akaike Information Criterion (AIC). 

\begin{figure*}
    \centering
    \includegraphics[width=0.9\textwidth]{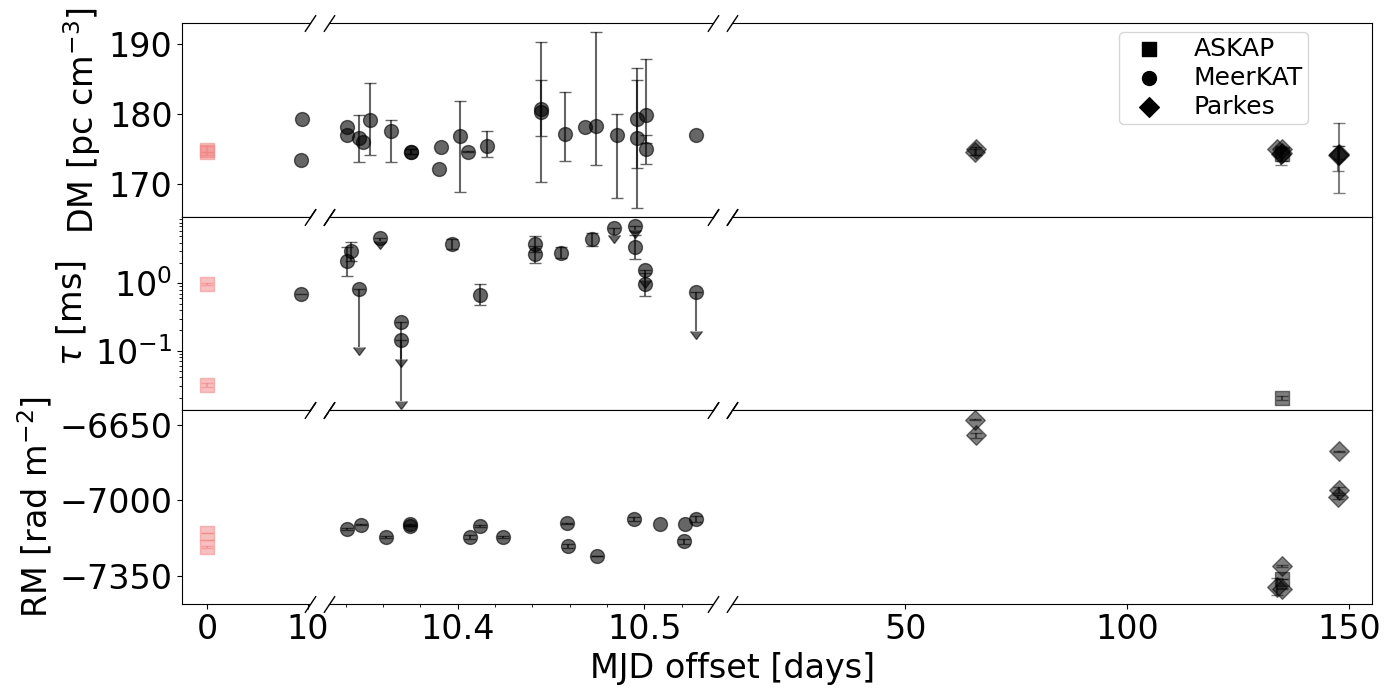}
    \caption{Top panel: DM evolution of FRB20250613. The Square, circle and diamond points show the ASKAP, MeerKAT and Parkes bursts respectively. The `red' points represent the discovery burst A1. Middle panel: burst $\tau$ at 1.0 GHz reference frequency assuming $\alpha = -4.0$. Bottom Panel: Burst RM. RM uncertainties are significantly smaller than the absolute RM values, making it difficult to see them. Each panel is separated by $x$-axis breaks into three parts. The middle part shows a zoom-in on the sample of MeerKAT bursts discovered on 24 June 2025. Time on the x axis is relative to the burst time of the first ASKAP burst, A1.}
    \label{fig:burst_properties}
\end{figure*}

For a handful of bursts, significant frequency modulations -- likely also the result of multi-path propagation -- were observed. Strong frequency modulations, when band-averaged, can cause changes in the apparent shape of the burst and complicate modelling of the scattering timescale using equation \ref{eq:scattering}. In these instances, a sub-band was manually selected and modelled instead. This method performs poorly when the S/N is low, which can lead to poorly constrained posteriors for complex models with multiple Gaussian pulses. For these faint bursts, we could only place upper limits on $\tau$, or in some cases, could not constrain the scattering timescale at all. The final scattering timescale measurements for each burst are surmised in Table \ref{tab:burst_tab1} and shown in the middle panel of Fig. \ref{fig:burst_properties}.

\subsubsection{Rotation measure and polarisation fraction analysis}
\label{sec: pol_analysis}

A time-based matched filter was applied to the Stokes \textit{I}, \textit{Q} and \textit{U} dynamic spectrum of each burst using the morphology fits outlined in Section \ref{sec: temp_analysis}. In cases where $\tau$ was unconstrained, a simple boxcar was used. Next, a mask was applied to filter out channels with a S/N < 3.0. To determine the rotation measure (RM), Stokes \textit{Q} and \textit{U} were normalised by Stokes \textit{I} to mitigate the strong frequency structures caused by multipath propagation. The normalised Stokes spectra were processed using the \code{RMtools} python package to fit the RM via RM synthesis \cite[][]{brentjens2005faraday}. The resulting RM, its uncertainty and power-averaged central frequency are recorded in Table~\ref{tab:burst_tab2}.

We used the derived RM to correct for Faraday rotation in the Stokes \textit{Q} and \textit{U} dynamic spectra by applying a rotation in \textit{Q}-\textit{U} space:

\begin{equation}
    \label{eq:FdayRot}
    \begin{split}
        & Q\mathrm{_{deRM}} = Q\mathrm{cos(2PA)} + U\mathrm{sin(2PA)} \\
        & U\mathrm{_{deRM}} = Q\mathrm{sin(2PA)} - U\mathrm{cos(2PA)}, \\
    \end{split}
\end{equation}

where $\mathrm{PA}$ is the polarisation position angle

\begin{equation}    
    \label{eq:RM}
    \mathrm{PA(\nu) = RMc^{2}}\bigg(\frac{1}{\nu^{2}} - \frac{1}{\nu_{0}^{2}}\bigg).
\end{equation}

Once the Faraday corrections were applied, we calculated the band-averaged, time-integrated polarisation fractions: 

\begin{equation}
\label{eq: pol_fracs}
\begin{split}
    & x = \mathrm{\frac{\sum X(f)}{\sum I(f)}} \\
    & L = \sqrt{Q^2 + U^2}, \\
    & P = \sqrt{Q^2 + U^2 + |V|^2}, \\
\end{split}
\end{equation}

where $|V|$ is calculated by integrating over $\mathrm{|V(f)|}$. To account for the bias introduced by taking the absolute value of $\mathrm{V(f)}$, we first de-biased the data following the methods of \cite{karastergiou2003v} and \cite{oswald2023pulsar1}. The polarisation fractions are recorded in Table \ref{tab:burst_tab2}.

\subsection{Optical observations and analysis}
\label{sec: host_galaxy}

\subsubsection{Imaging and Host Association} \label{sec:host-imaging}

The host galaxy was first identified in DESI Legacy Survey DR10 \citep{DECALS} $r$-band imaging at RA 04\hmshh36\hmsmm33.06\hmsss\ and DEC -44\degree31\arcm58.92\arcs. Using the DESI data, a Probabilistic Association of Transients to their Hosts (PATH; \citealt{path}) analysis favours this galaxy as the host at $P(O\vert x) = 0.954$ assuming $P(U) = 0.2$ and standard PATH priors, representing a robust host association. 

\begin{figure}
    \centering
    \includegraphics[width=\linewidth]{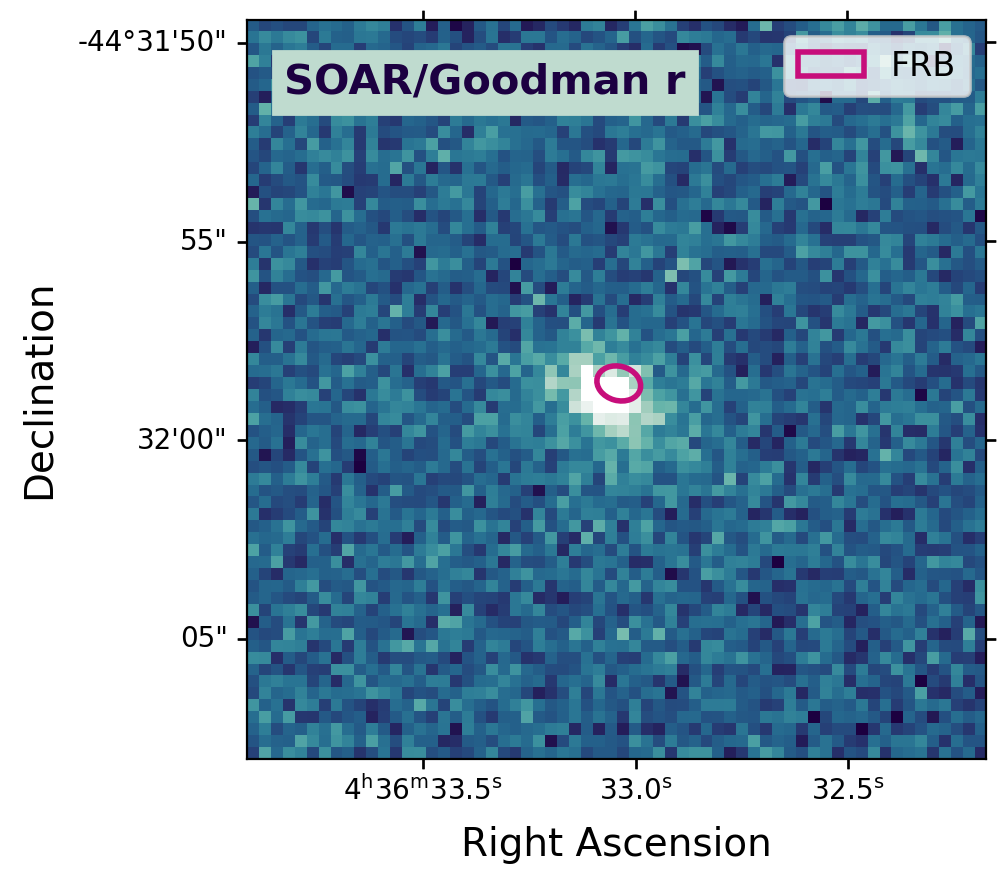}
    \caption{SOAR/Goodman $r$-band image of the host of FRB\,20250613A. The 1$\sigma$ FRB localisation is shown as a magenta ellipse.}
    \label{fig:host-image}
\end{figure}

We obtained further $r$-band imaging with the 4-m Southern Astrophysical Research Telescope (SOAR; Program ID SOAR2025B-031, PI Gordon) with the Goodman spectrograph \citep{SOAR} on 2026 January 7 UTC for 12x300s of exposure (Fig.~\ref{fig:host-image}). The data were reduced with the \texttt{photpipe} pipeline \citep{Rest+05} which applies flat-field and bias corrections, solves the WCS with reference to Gaia DR3, performs sky-subtraction and stacking with \texttt{swarp}, and flux calibrates with respect to SkyMapper DR2. We performed aperture photometry using a custom script\footnote{https://github.com/charliekilpatrick/photometry} and measure an $r$-band magnitude of 22.14 $\pm$ 0.08 AB, corrected for Galactic extinction following the \citet{Fitzpatrick:2007} extinction law. While the host association is already robust from the DESI imaging, a PATH analysis on the SOAR imaging returns an unambiguous $P(O\vert x) = 0.999$ association due to the superior imaging depth.

\subsubsection{Gemini Spectroscopy} \label{sec:host-spectroscopy}

To identify the host redshift, we obtained a spectrum with the Gemini Multi-Object Spectrograph (GMOS; \citealt{Hook-GMOS,Gimeno-GMOS}) on the 8-m Gemini South telescope (Program ID GS-2025B-FT-103, PI Gordon) via a Fast-Turnaround program. GMOS was configured using the B480 grating with central wavelengths at 640\,nm and 650\,nm, $2\times2$ binning, and a slit width of 1 arcsecond. We obtained $6\times900$\,s of exposure across 2025 September 16, October 27, and October 29 UTC. The data were reduced with the Python Spectroscopic Data Reduction Pipeline (\texttt{PypeIt}; \citealt{pypeit:zenodo,pypeit:joss_pub}) which performs bias-subtraction, flat-fielding, cosmic ray masking, and wavelength calibrations on each frame. After identifying and extracting the one-dimensional spectra of the host, we performed one-dimensional coaddition, flux calibration with respect to a spectrophotometric standard star, and telluric correction. We identify H$\beta$, [O\,{\sc iii}]$\lambda\lambda 4959, 5007$, H$\alpha$, and [S\,{\sc ii}]$\lambda\lambda 6716, 6731$ all in emission at $z = 0.0987 \pm 0.0001$ which we adopt as the redshift of FRB\,20250613A.

\subsubsection{Galaxy Light Profile Modelling and Offset Analysis} \label{sec:galfit-offset}

We use the light profile modelling software \texttt{Galfit} \citep{galfit_2002,galfit_2010} to model the surface brightness profile of the host galaxy using the SOAR imaging. We initialise the \texttt{Galfit} model using the image point spread function derived via \texttt{photutils} \citep{photutils} \texttt{EPSFBuilder}, a single standard S\'ersic profile \citep{Sersic_1968}, a sky background model, and initial guesses on the galaxy's central position, apparent magnitude, effective radius $r_e$, S\'ersic index $n$, semi-minor/semi-major axis ratio ($b/a$), and position angle. The host is best described by a S\'ersic profile with $n=0.75$, $b/a = 0.66$, and $r_e = 1.40$\arcs\ (2.58~kpc assuming WMAP9 cosmology; \citealt{WMAP9}) which is consistent with a slightly inclined disk galaxy.

To quantify the location of the FRB within its host, we measure the projected galactocentric offset using a custom script\footnote{https://github.com/muethela/little-astro-scripts} accounting for the astrometric tie uncertainty of the image and the host positional uncertainty. Assuming the FRB localisation is described by a two-dimensional Gaussian probability map, the script calculates the separation between the host centre and every pixel within the 5$\sigma$ localisation ellipse to build a distribution of angular offsets weighted by the probability map. From this distribution we derive the median and 68 per cent CI of the projected angular offset, which for FRB\,20250613A is $0.60^{+0.37}_{-0.26}$\arcs. We further convert to physical ($1.11^{+0.69}_{-0.47}$~kpc) and host-normalised ($0.43^{+0.27}_{-0.18}~r_e$) offsets assuming WMAP9 cosmology and using the effective radius derived from \texttt{Galfit}, respectively. While offset measurements are not available for every localised FRB, when compared to the
 sample with available measurements derived by \citet{Gordon+25}, FRB\,20250613A is closer to its host galaxy centre than the bulk of the FRB population, which have a median host-normalised offset of 1.05~$r_e$.

\subsubsection{Spectral Energy Distribution Analysis} \label{sec:SED}

To derive the host stellar population properties, we use the Bayesian inference framework \texttt{Prospector} \citep{Johnson+21}, jointly fitting the host photometry and spectroscopy. In addition to the SOAR $r$-band photometry, we compile $g$-, $r$-, and $z$- band photometry from DESI Legacy Survey DR10 and $i$- and $Y$-band photometry from the Dark Energy Survey (DES; \citealt{DES}). All photometry and spectroscopy are corrected for Galactic extinction following the \citet{Fitzpatrick:2007} extinction law. To initialise the \texttt{Prospector} model,  we use the stellar population synthesis library \texttt{python-fsps} \citep{Conroy2009, Conroy2010}, employing an 8-bin continuity prior non-parametric star-formation history \citep{Leja2019}, a nebular emission model which includes marginalisation over the emission lines, a pixel outlier model in photometry and spectroscopy, a spectral smoothing model, and model to normalise the spectroscopy to the photometry using a 12th order Chebyshev polynomial. We additionally assume a \citet{Kroupa01} initial mass function, \citet{KriekandConroy13} dust attenuation curve, and require adherence to the \citet{Gallazzi2005} mass-metallicity relation. For further details on the priors and their allowed ranges, refer to \citet{Gordon+23}. Once initialised, the posteriors are then sampled using the dynamic nested sampling routine \texttt{dynesty}.

\begin{figure*}
    \centering
    \includegraphics[width=0.52\textwidth]{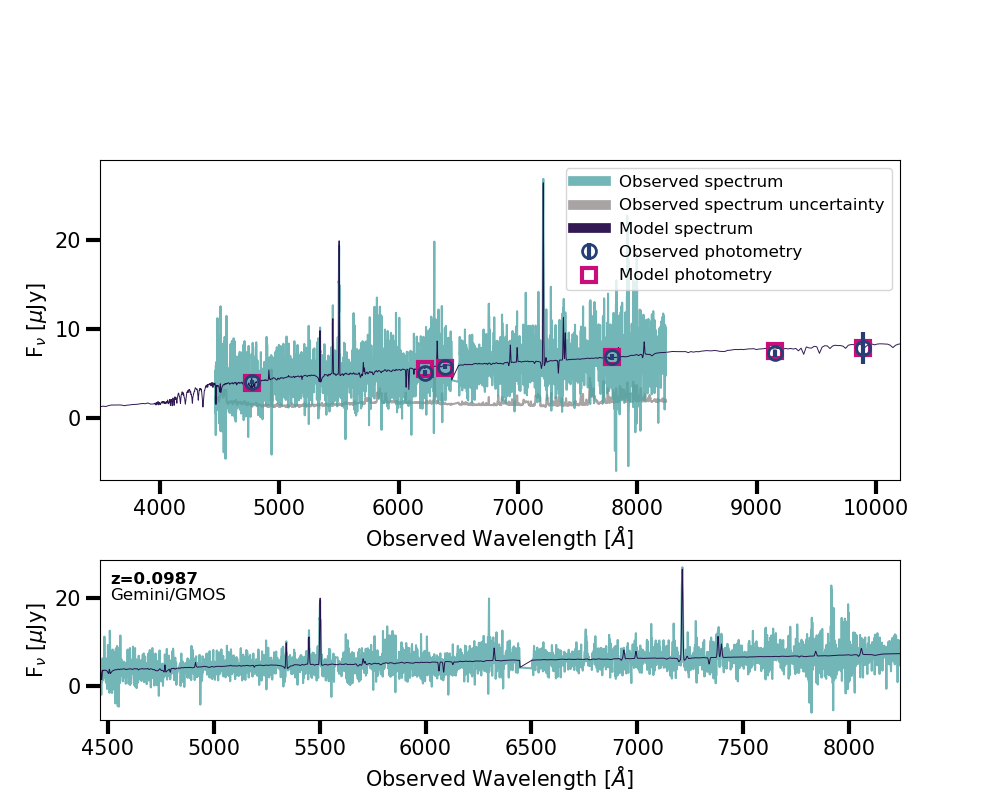}
    \includegraphics[width=0.45\textwidth]{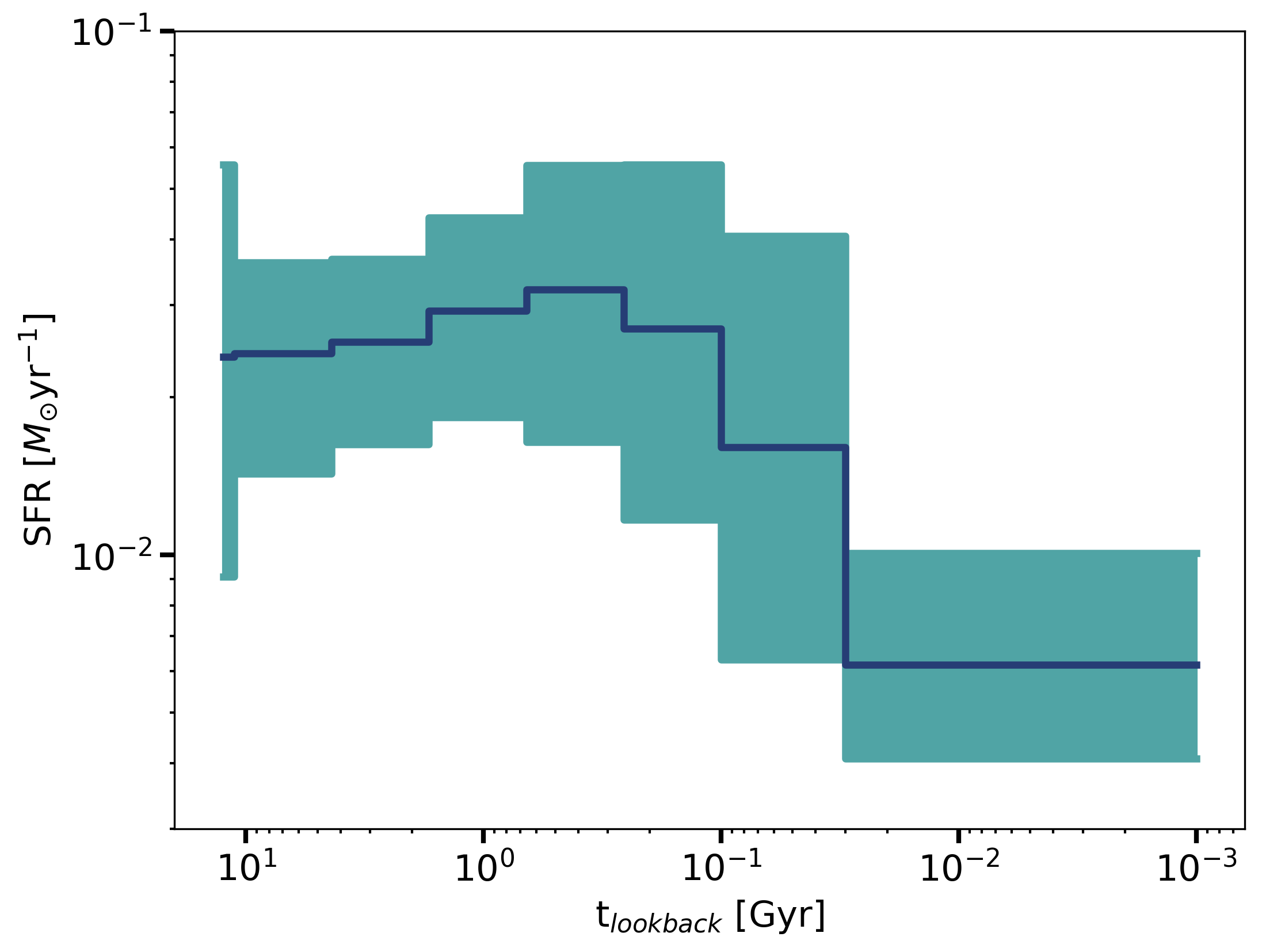}
    \caption{Left: Spectral energy distribution of the host of FRB\,20250613A, jointly fitting the photometry and spectroscopy. A zoom-in of the observed and modelled spectrum is shown in the bottom panel. Right: Star formation history of the host of FRB\,20250613A. The solid blue line represents the median star formation rate and the shaded teal regions are the corresponding 68 per cent CI.}
    \label{fig:SED-SFH}
\end{figure*}

The best-fit model, as determined by a combination of the model evidence statistic and visual inspection of the spectral energy distribution (SED, Fig.~\ref{fig:SED-SFH}), reveals the host of FRB\,20250613A is a moderately old (the mass weighted age is $t_{\rm m} = 5.96^{+1.09}_{-1.25}$ Gyr), low-luminosity (1.63$\times10^{8} L\odot$) dwarf galaxy with a stellar mass of log(M$_*$/M$_{\odot}$)$ = 8.31 \pm 0.09$ and a sub-solar gas-phase metallicity of log(Z$_{\rm gas}$/Z$_{\odot}$)$ = -0.65^{+0.23}_{-0.25}$ (all values reported as median and 68 per cent CI). Despite the low integrated 0-100 Myr star-formation rate (${\rm SFR}_{\rm 0-100 Myr} = 0.01^{+0.02}_{-0.01}\,M_{\odot}$~yr$^{-1}$), the host is still considered actively star-forming per the \citet{Tacchella+22} mass-doubling number criterion, which depends on the specific star-formation rate (sSFR = SFR/M$_*$), with log(sSFR [yr$^{-1}$])$ = -9.90^{+0.34}_{-0.40}$. The star-forming designation is further corroborated by the host's location on a Baldwin-Phillips-Terlevich (BPT; \citealt{BPT}) diagram. The star-formation history is characterised by a relatively constant SFR across the host's evolution with a recent quenching event in the last 30~Myr, as shown in Fig.~\ref{fig:SED-SFH}.

To place the host of FRB\,20250613A in the context of the known host population with modelled stellar population properties, we compare with the results of \citet{Gordon+23}. We find FRB\,20250613A has a similar sSFR and mass-weighted age to the bulk of the full host population but also a considerably lower stellar mass than even most repeaters (which tend to occur in lower-mass galaxies than apparent non-repeaters, although for a counterpoint see \citealt{Muller+25} who present an apparent non-repeater in a very low-mass dwarf host). The host of FRB\,20250613A bears a striking resemblance to the host of the repeating FRB\,20121102A, with a similar stellar mass, SFR, mass-weighted age, gas-phase metallicity, and luminosity.

\section{Properties of FRB20250613A}
\label{sec:Methods part2}

\subsection{Variations in scattering}
\label{sec:scatter_var}
There are large apparent variations in scattering observed on timescales of minutes to hours between bursts during the MeerKAT epoch on 24 June that spans almost two orders of magnitude, with the smallest and largest scattering timescales being 0.14$\pm$0.18 ms and 7.2$\pm$1.0 ms (referenced to a frequency of 1.0 GHz assuming a spectral index of -4) for bursts M6B and M14, respectively. Several other repeating FRBs have previously shown evidence for variations in temporal broadening due to multi-path propagation. For example, \cite{ocker2023scattering} detected statistically significant variations in scattering times (of at least a factor of $\sim$2) in bursts from FRB 20190520B. \cite{panda2025low} also reported large variations in best-fit scattering time between bursts from FRB 20240114A on timescales of hours using GMRT, but without reporting the statistical significance.

While statistically we have very clear evidence of scattering time variability in FRB~20250513A, the scattering times have been estimated solely based on modelling the temporal features of the burst. To verify that the broadening structure is indeed consistent with multi-path propagation, we investigated the frequency dependence of the inferred scattering tail:

\begin{equation}
    \label{eq: specindex}
    \tau(f) = \tau_{c} \bigg(\frac{f}{f_{c}}\bigg)^{\alpha},
\end{equation}

where $\tau_{c}$ is the scattering timescale at the central frequency $f_{c}$ and $\alpha$ is the scattering index. A steep frequency dependence in $\tau$ is a strong indication for the presence of multipath scattering. For example, Kolmogorov plasma turbulence-which is common in the ISM- is expected to have a spectral index of $\alpha = -4.4$ \citep{lorimer2005handbook}. Deviations from the Kolmogorov spectrum are both theoretically expected (due to, e.g., the presence of inner and outer scales to the turbulence) and observed \citep{geyer2017scattering}. But chromaticity -- with a steep spectrum -- is ubiquitous.

\begin{figure*}
    \centering
    \includegraphics[width=0.8\textwidth]{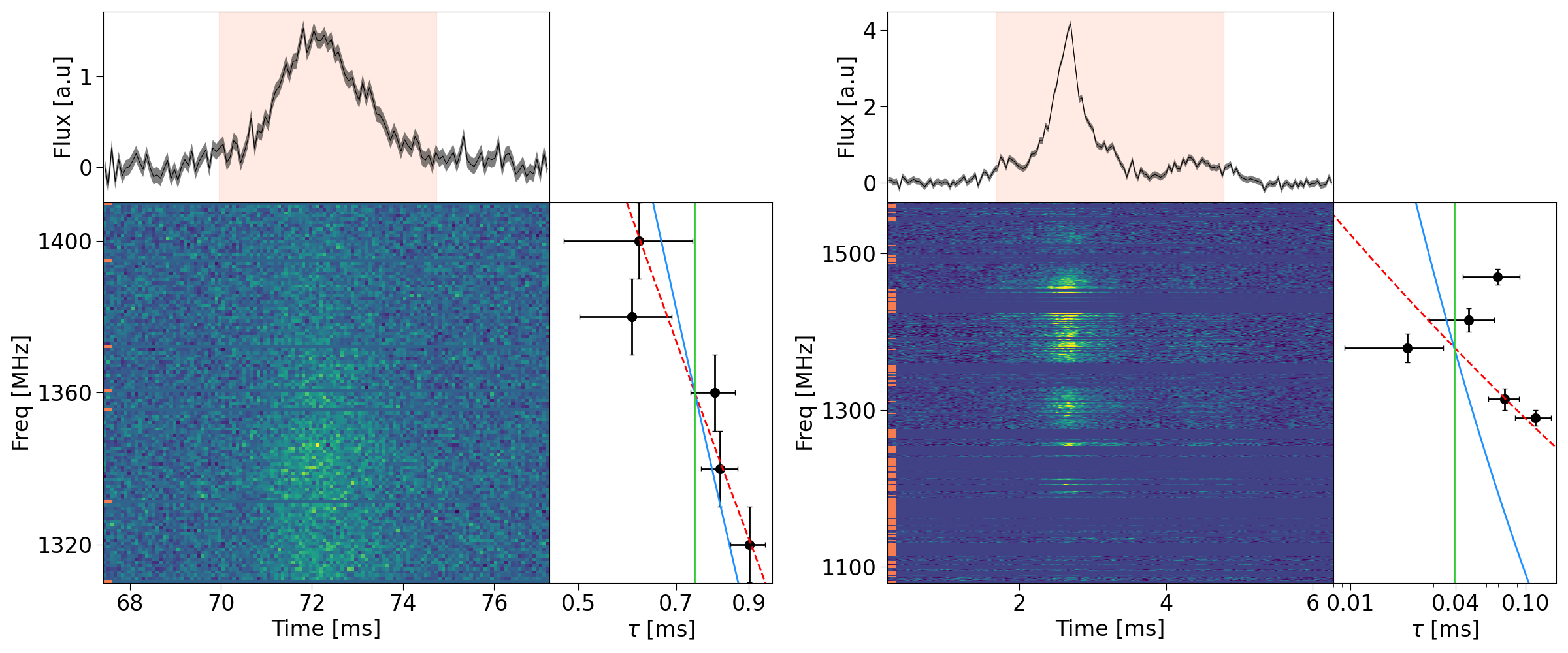}
    \caption{Results of $\alpha$ fitting. Left: Burst M11. The top panel shows the time-series with the on-pulse region highlighted in red. The middle panel shows the dynamic spectrum. Channels that are flagged are highlighted in the orange patches on the left side of the spectrum. The panel to the right of the dynamic spectrum shows the sub-banded scattering index fit of $\alpha$ = -6.5 $\pm$ 1.7. The red line is the least squares fit whilst the blue and green lines show what a -4.0 and 0.0 value for $\alpha$ would look like. The error bars along the y-axis indicate the bandwidth extent for each sub-band. Right side: M6B burst with a fitted scattering index of $\alpha$ = -13.8 $\pm$ 8.1.}
    \label{fig: meerkat_scattplot}
\end{figure*}

We analysed two bursts from the MeerKAT sample, M11 and M6B which are shown in Fig. \ref{fig: meerkat_scattplot}. Although M14 possesses the largest inferred scattering timescale of 7.2$\pm$1.0 ms, the burst is confined to a narrow spectral bandwidth, making it difficult to investigate the degree of chromaticity of the pulse broadening. To measure the spectral index confidently, we chose M11 since it still has a relatively large scattering index of $\tau$ = 2.9 $\pm$ 0.6 ms but has a wider spectral occupancy than M14. To fit the scattering index, the Stokes \textit{I} spectrum was split into multiple sub-bands, each of width $\Delta f \geq$ 20 MHz, and frequency averaged. Sub-bands were not necessarily contiguous due to the presence of strong scintillation features; their central frequencies were chosen manually to maximise signal--to--noise ratio within each sub-band. 

We employed a conservative method for fitting the scattering index. First, the frequency averaged profile of each sub-band was modelled as a sum of $\mathrm{N}$ Gaussian pulses, where $\mathrm{N}$ was the number of pulses that best-fit that the full band-averaged data supported, using the criteria described in Section \ref{sec: temp_analysis}. For each Gaussian pulse, the width was set to a constant value (which was derived in Section \ref{sec: temp_analysis}). We sampled the amplitude of each pulse across each sub-band to ensure any spectral features were taken into account. Similarly, we sampled the position of the leading Gaussian pulse, whilst keeping the displacement between each neighboring pulse fixed, to identify any residual dispersion hidden in the burst. We observed a relatively large drift in the pulse position of burst M11, which we estimated as a differential dispersion delay of $\delta$DM = -1.776 pc cm$^{-3}$ compared to the mean value used initially. We incoherently de-dispersed the Stokes dynamic spectra of burst M11 before repeating both the full-band and sub-band fits to obtain the spectral index; the scattering timescale of $\tau$ = 2.9\,ms was obtained after applying this dispersion correction. A scattering index of $\alpha$ = -6.5 $\pm$ 1.63 and -13.8 $\pm$ 8.1 were obtained for M11 and M6B, respectively. The fits of each burst are shown in Fig.~\ref{fig: meerkat_scattplot}. 

\begin{figure}
    \centering
    \subfigure[]{\includegraphics[width=0.8\linewidth]{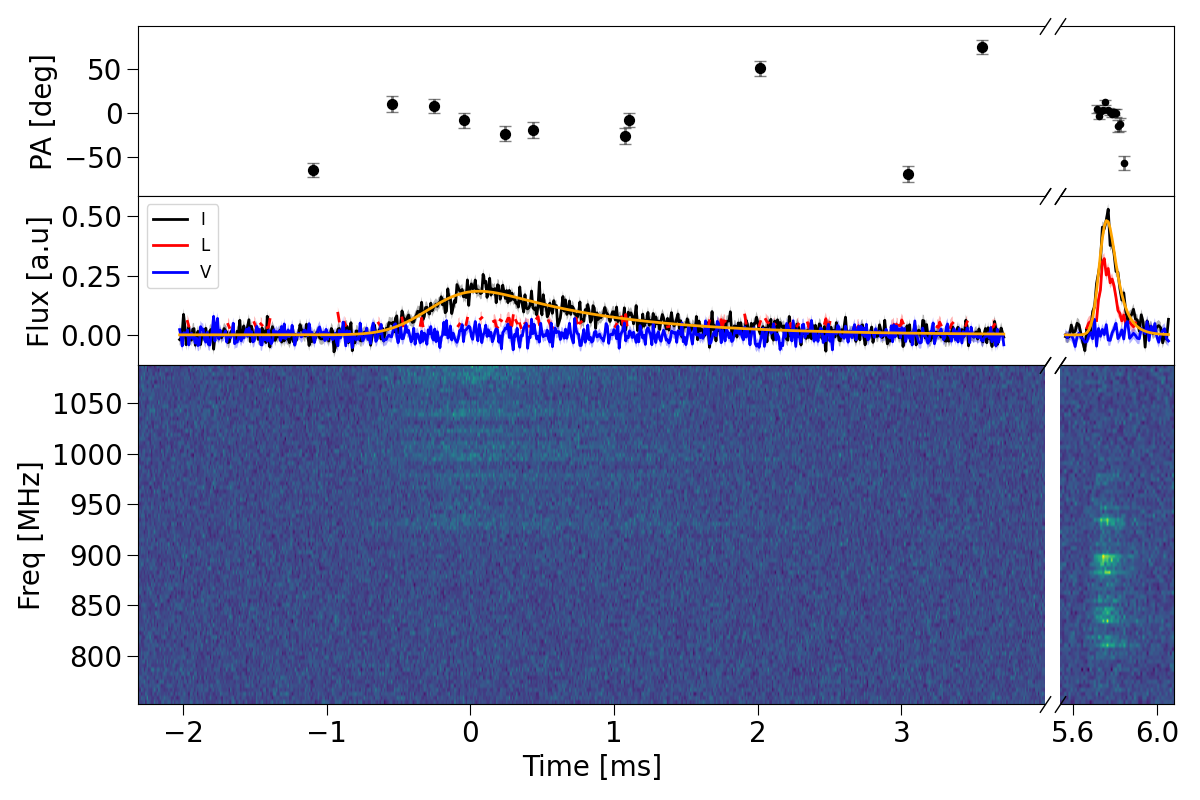}}
    \subfigure[]{\includegraphics[width=0.8\linewidth]{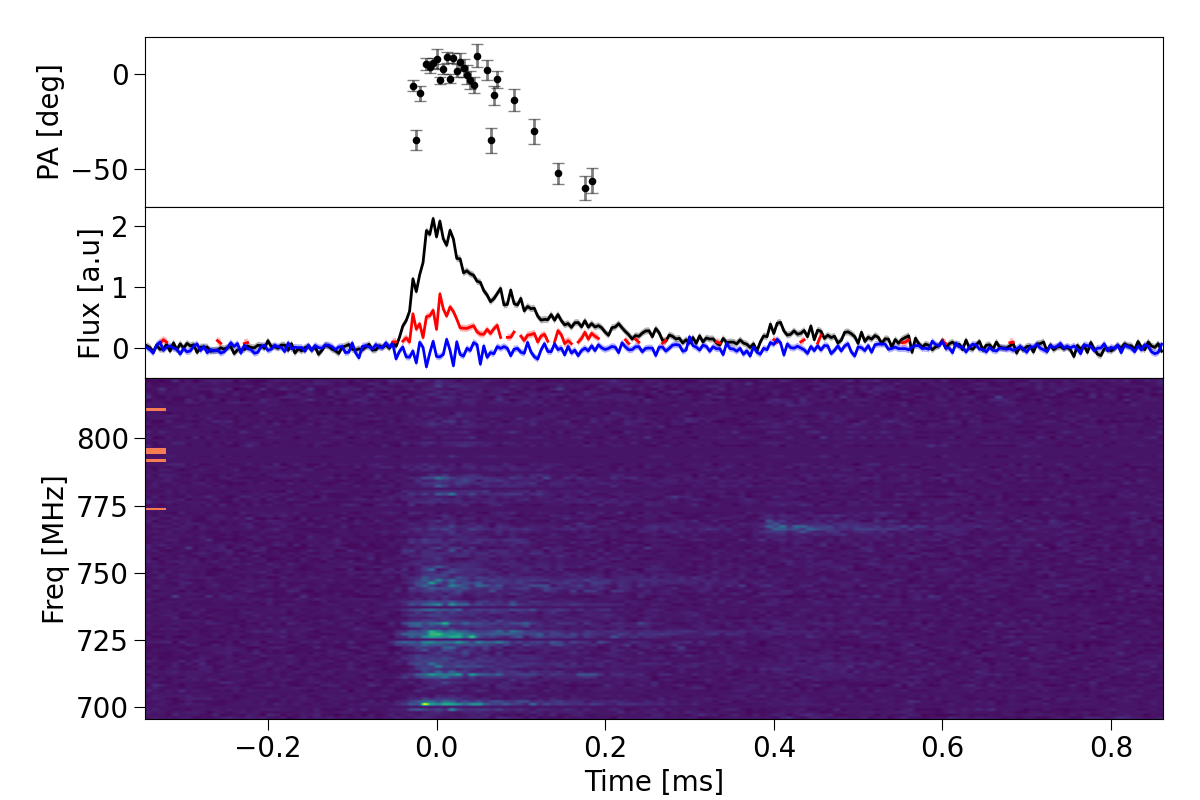}}
    \subfigure[]{\includegraphics[width=0.8\linewidth]{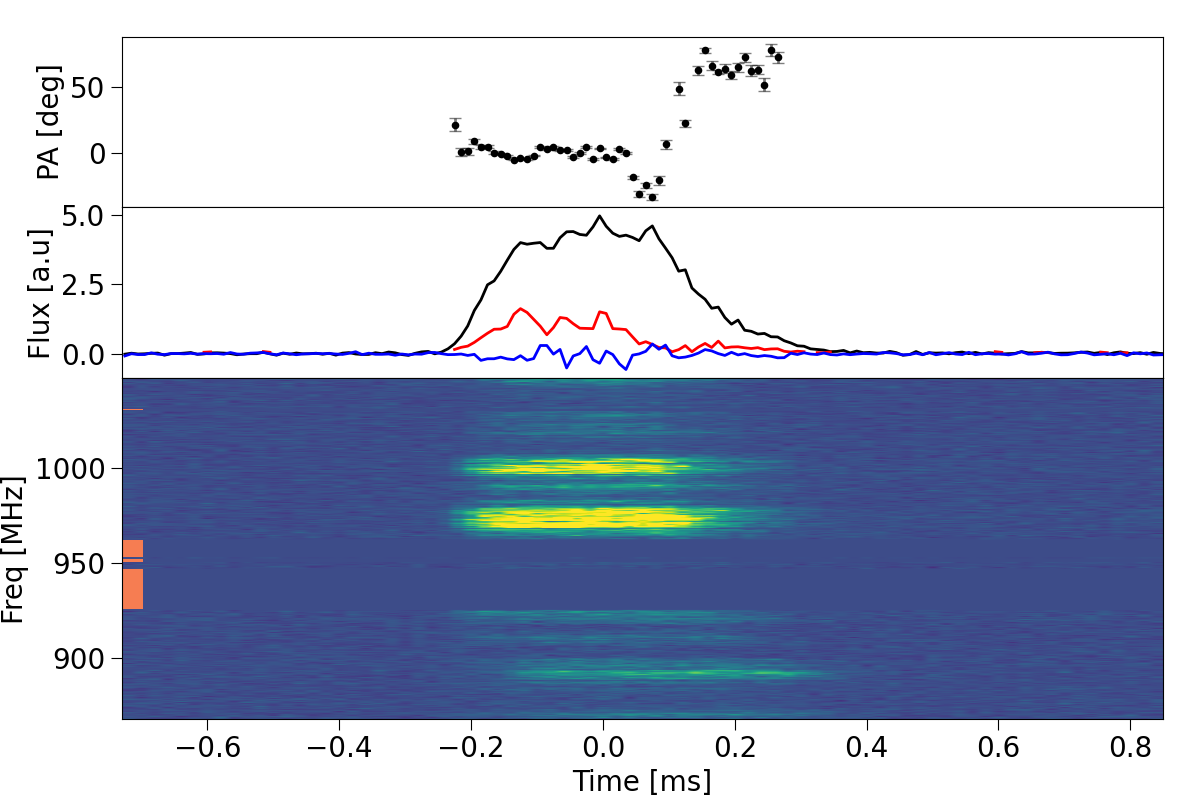}}
    \caption{(a) The ASKAP discovery burst A1. Top panel: Polarisation Position Angle (PA) profile. Middle panel: Time series \textit{I}, \textit{L} and \textit{V}. Bottom panel: Stokes \textit{I} dynamic spectrum. The break in $x$-axis is used to indicate that each component has been analysed separately with a different DM. Data on the leftmost panels of the break are presented at a time resolution of 20 \textmu s and data on the rightmost panels of the break are at a resolution of 10 \textmu s. (b) Burst A2, presented at a resolution of 2 \textmu s. Flagged channels are represented with an orange band along the left side of the dynamic spectrum. (c) The leading component of burst M24 presented at a  resolution of 2 \textmu s.}
    \label{fig:ASKAP_pol_mosaic}
\end{figure}

As shown on the left panel of Fig. \ref{fig: meerkat_scattplot}, burst M11 demonstrates a strong negative spectral dependence in scattering time measurements. While the uncertainties on the subbanded scattering time measurements are large (due to wide intrinsic pulse width, and -- especially at the top of the band -- the lower flux density), the fitted scattering times are consistent with the expected degree of chromaticity. In contrast, while the temporal broadening of burst M6B is clearly frequency dependent, it is not well fit by the power law in equation \ref{eq: specindex}. The inferred scattering time is extremely small near the centre of the band, with $\tau$ = 0.02 ms. The sharp decline in flux after the peak places strong upper limits on the scattering times at all frequency ranges, and we infer that unmodelled burst substructure likely causes the increase in fitted scattering time at the top end of the band. In essence, the very small scattering time (relative to the width of the intrinsic sub-burst structure) makes it difficult to accurately measure the scattering time, even though it is easy to constrain it to be small. Given these measurements and caveats, we are reasonably confident that both bursts display scattering as a result of multi-path propagation and accordingly the source does exhibit large variations in scattering over timescales of minutes to hours.

\begin{figure*}
    \centering
    \includegraphics[width=0.8\textwidth]{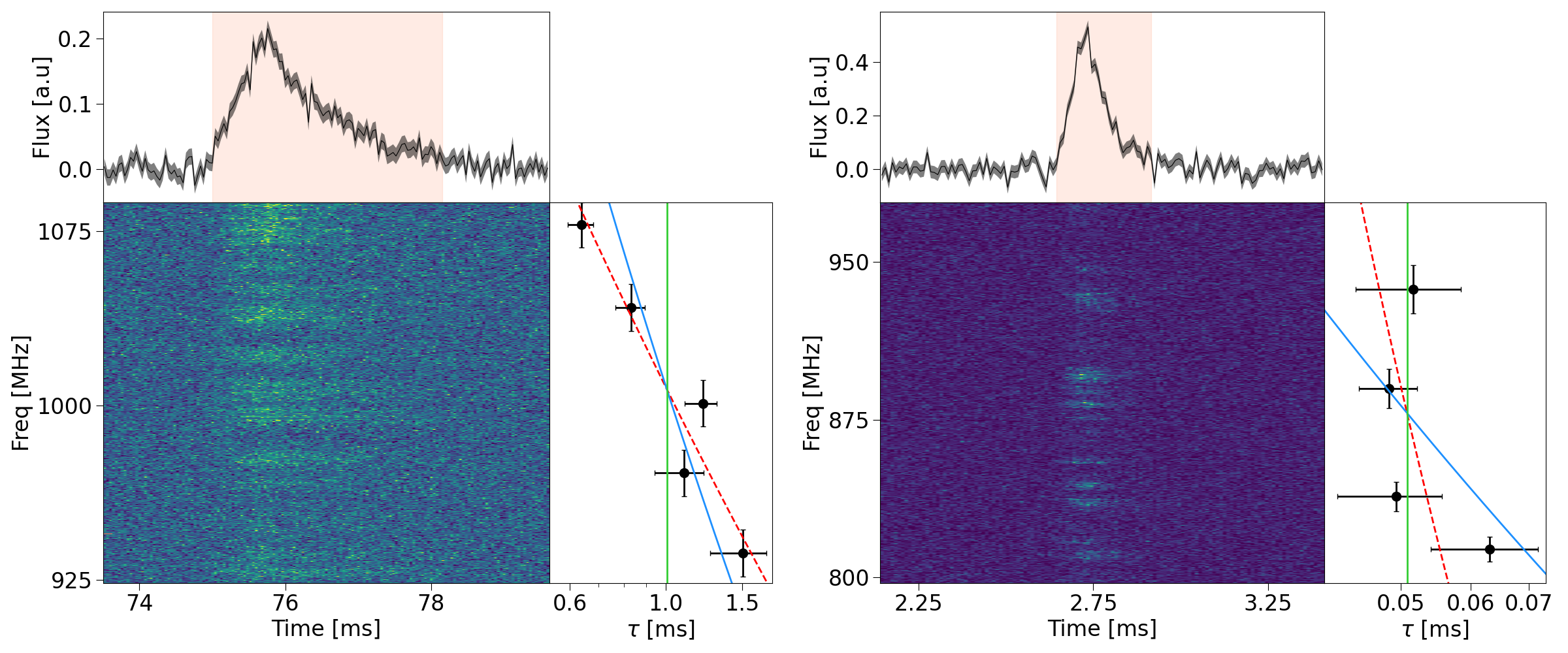}
    \caption{Results of $\alpha$ fitting for the A1 burst. Left side: Component 1 (A1A), $\alpha$ = -6.2 $\pm$ 1.1. Right side: Component 2 (A1B), $\alpha$ = -1.1 $\pm$ 1.3.}
    \label{fig: askap_scattplot}
\end{figure*}

We also observe apparent variations in scattering on much shorter timescales of $\lesssim10$ milliseconds. As shown on the left panel of Fig. \ref{fig:ASKAP_pol_mosaic}, burst A1 possesses two components which, when isolated and structure maximised, are each best fit with a single scatter broadened Gaussian pulse. The fitted scattering timescales of $\tau$ = 0.97$\pm$0.04 and 0.031$\pm$0.017 ms respectively display a variation by more than a factor of 30. As shown in the left panel of Fig. \ref{fig: askap_scattplot} the leading (broad) component has a steep slope of $\alpha = -6.2 \pm 1.1$ that is well fit by a power law function. For the second (narrow) component, shown in the right panel of Fig. \ref{fig: askap_scattplot}, the measured scattering times are much lower despite the lower central frequency. Furthermore, the scattering time does not change towards the top of the band. As with burst M6B, this is likely related to residual unmodelled burst structure as the scattering time becomes very small compared to the modelled burst width. This large change in scattering over a duration of just $\sim$6-7 ms demands a different interpretation than the variations we see on much longer timescales, which we explore in Section \ref{sec:nonlinear}.

There are caveats to using a one-sided exponential scattering tail as defined in Equation \ref{eq:scattering} as the pulse broadening function (PBF). This PBF assumes the projected scattering disk of the source is Gaussian in shape. However, scattering disks that do not satisfy this assumption can decay more slowly at large time delays than the exponential function predicts, and indeed there is substantial observational evidence for such configurations. \cite{geiger2025nanograv} show that incorrect assumptions in the underlying PBF can introduce systematic biases in the measured scattering times and frequency dependence. We cannot exclude this impacting our measured values, meaning the uncertainties quoted may be somewhat underestimated. However, this effect cannot explain the order of magnitude (and greater) changes in scattering that we observe between bursts.

\subsection{Constraints on scattering screen distance}

Diffractive scintillation results in both temporal broadening and spectral modulation, characterised by the scattering time $\tau$ and the scintillation bandwidth $\nu_{dc}$. These two parameters are related as follows:

\begin{equation}
    C = 2\pi\tau\nu_{dc},
\end{equation}

where the scattering constant $C$ is roughly unity for a single scattering screen \citep{lambert1999theory}. FRB sources often show a scattering constant $\gg$1.0 \citep{sammons2023two}, implying the scintillation (seen in the frequency domain) and scattering (seen in the time domain) are caused by two different screens. If a Galactic screen produces the scintillation features present in the burst, another screen closer to the source must be responsible for the observed scattering. The condition for strong scintillation is that the source (as angularly broadened by the first screen) must remain unresolved at the Galactic screen, which enables upper limit constraints on the two-screen geometry:

\begin{equation}
    L_{x}L_{g} \lesssim 1.6\times10^{-4}\frac{D_{s}^{2}}{ \nu^{2}_{c}(1+z)} \frac{\nu_{dc}}{\tau}
    \label{eq:twoscreen},
\end{equation}

where $L_{x}$ is the distance between the source and first scattering screen,  $L_{g}$ is the distance between the observer and Galactic screen, $D_s$ is the luminosity distance in kpc, $\tau$ is the scattering time in milliseconds, and $\nu_{c}$ and $\nu_{dc}$ are the central frequency and scintillation bandwidth in MHz, respectively. We measured the scintillation bandwidth of the burst sub-component A1B by modelling the autocorrelation of the spectrum as a Lorentzian:

\begin{equation}
    \mathrm{ACF}(\Delta\nu) = m^2\frac{\nu_{dc}^2}{\nu_{dc}^{2} + \Delta\nu^{2}},
\end{equation}

where $m$ is the modulation index and $\Delta\nu$ is the frequency lag. We calculated the ACF of the residual spectrum by first subtracting away the broad frequency structure of the burst which was modelled using a third order polynomial. We obtained a scintillation bandwidth of 1.46 $\pm$ 0.06 MHz and a modulation index of 0.83 $\pm$ 0.02. Using the updated NE2025 galactic electron density model \citep{ne2025} the expected scintillation bandwidth from the Milky Way is 0.995 MHz. Hence we conclude that the scintillation features are consistent with a Galactic screen. 

The modulation index $m$ is $\sim$0.83, close to the value of unity expected if the scatter-broadened source remains unresolved by the galactic screen \citep{pradeep2025scintillometry}. If the scatter broadened source is resolved by the Galactic screen, we would expect a smaller modulation index, which additionally is expected to decrease across the burst \citep{sammons2023two, pradeep2025scintillometry}. We made attempts to measure $m$ across the burst but were unsuccessful because of the low S/N. Therefore, while our observations are consistent with an unresolved source screen, we cannot rule out a very mildly resolved screen.

Assuming the screen is unresolved, we can place constraints on the screen geometry. Using a $\Lambda$-CDM cosmology with Hubble constant of $H_{\mathrm{0}}$ 67.4 km/s/Mpc and a matter density of $\Omega_{m}$ = 0.315 to estimate the luminosity distance \citep{aghanim2020planck}, a scintillation bandwidth of 1.0 MHz and a maximum scattering time of 7.2 ms at 1.0 GHz, we obtain an upper limit on $L_xL_g \lesssim$ 6.55 $\pm$ 1.8 kpc$^{2}$. If we assume the Galactic screen lies at a distance given by the scale height of the Milky Way thick disk \cite[1.57 $\pm$ 0.15 kpc, ][]{ocker2020electron}, we would place an upper limit of 4.17 $\pm$ 0.85 kpc of $L_x$. However, since the true location of the Galactic screen is unknown, this upper limit is not particularly tight; if the screen is located closer to the Earth, the upper limit would increase proportionally. In any case, the scintillation measurements strongly favour a screen within the host galaxy, but do not constrain it to be in the immediate vicinity of the source.

\subsection{Variations in DM and RM}

We see apparent variations of both DM and RM on short timescales of minutes to hours. While formally significant, the variation in DM on the order of a few pc cm$^{-3}$ should be taken with some caution, due to the difficulty in measuring DM for individual bursts via structure maximisation.  There is, however, good support for RM varying on timescales of minutes to hours (by $\sim$20-50 rad m$^{-2}$), as seen in Fig. \ref{fig:burst_properties}. RM changes on longer monthly timescales are even more significant, showing large variations of $\Delta$RM $\sim$300 rad m$^{-2}$ around a large mean RM of $-$7134 rad m$^{-2}$. Conversely, there is no evidence for any variation in the mean DM away from a constant value of $\sim$174 pc cm$^{3}$ over these same month-long timescales. Put together, this demands (as expected) that a highly magneto-ionised region near the source contributes most of the RM but only a small fraction of the DM.

\subsection{Depolarisation}

An increasing number of repeating FRB sources show evidence of being embedded in highly turbulent and highly magnetised ionised environments \citep{lu2023temporal, mckinven2023large}. A far smaller fraction of apparently non-repeating FRBs have shown similar frequency-dependent depolarisation, suggesting that these sources typically reside in less complex magneto-ionic environments, although the measurement of depolarisation for non-repeating FRBs is hindered by the relatively limited bandwidth of FRB search systems \citep{uttarkar2026depolarization}.
In any case, bursts that undergo multipath-propagation through a dense and magnetised turbulent ionised medium will -- in a process known as spectral depolarisation (Burn's depolarisation; \citealp{burn1966depolarization}) -- see an exponential suppression in the total linear polarisation at longer wavelengths \citep{amiri2021first, feng2022frequency, lu2023temporal}:

\begin{equation}
\label{eq: burnslaw}
    p_{m}(f) = p \cdot \mathrm{exp(-2\sigma_{RM}^{2}\lambda^4)},
\end{equation}

where $p_{m}$ and $p$ are the measured and intrinsic total polarisation fractions, respectively, $\mathrm{\lambda}$ is the wavelength and $\mathrm{\sigma_{RM}}$ is the RM scatter that represents the strength of the magneto-ionic complexity. The degree of RM variability along the multiple paths traversed by the bursts will depend on both the transverse gradients in RM and the size of the deflections. Therefore, $\mathrm{\sigma_{RM}}$  is expected to correlate with both $|\mathrm{RM}|$ and $\tau$. These relationships have been measured empirically by \cite{feng2022frequency} using a power-law model and a small sample of repeating FRBs, showing a best fit in which $\mathrm{log\sigma_{RM}} \propto 0.81(16)\mathrm{log}\,\tau$ and $\mathrm{log\,\sigma_{RM}} \propto 0.62(30)\mathrm{log}\mathrm{|RM|}$, where $\mathrm{\sigma_{RM}}$ and $\mathrm{|RM|}$ are in units of rad m$^{-2}$ and $\tau$ is in ms. Variations in $\tau$ and $\mathrm{|RM|}$ that have been observed in some repeaters have been evaluated for consistency with this $\mathrm{log\sigma_{RM}}$ framework. For instance, temporal variations in both $\mathrm{\sigma_{RM}}$ and $\mathrm{RM}$ were observed in FRB~20201124A \citep{lu2023temporal}, which were broadly consistent with \cite{feng2022frequency} and supported by a single complex magneto-ionised environment local to the FRB source; temporal variations in $\tau$ were not explored in \citep{lu2023temporal}.

FRB~20250613A, with a moderately large sample of bursts displaying apparent time variability in RM and pulse broadening, offers an additional opportunity for testing this framework. We analysed a subset of bursts from the full sample that had large bandwidths and were not excessively modulated by scintillation such that we could divide them up into at least 3 sub-bands with a good balance of bandwidth and S/N. The bursts M6B, M9, M24, A1B and P2 were chosen and divided into sub-bands manually to measure their total polarisation fractions. Finally, for each burst we fitted $\mathrm{\sigma_{RM}}$ and $p$ using equation \ref{eq: burnslaw}. For comparison we also fitted each burst using the un-modified Burns law by assuming $p$ to be unity. 

\begin{figure}
    \centering
    \includegraphics[width=0.9\linewidth]{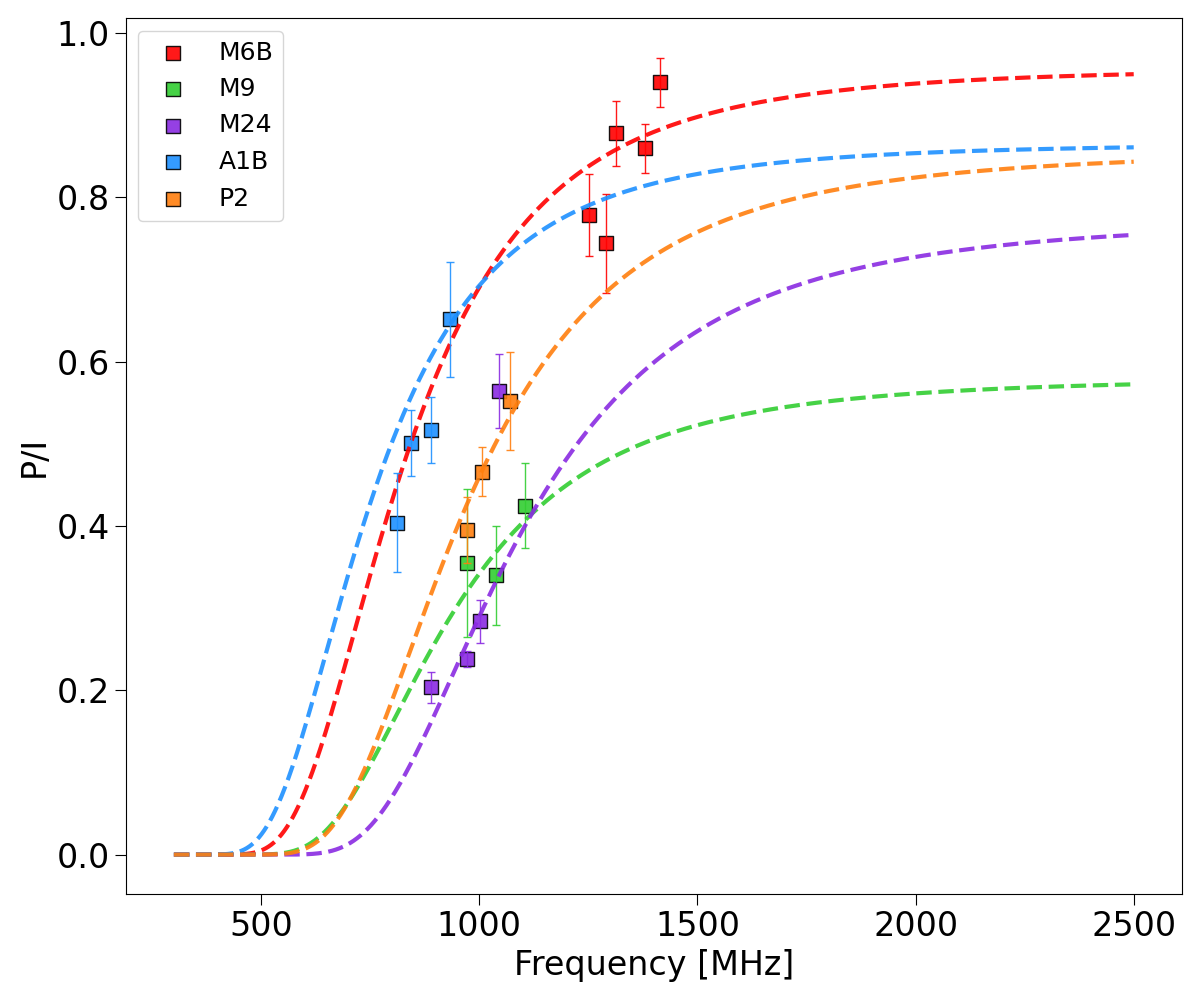}
    \caption{Burns depolarisation fitting for a sub-sample of MeerKAT bursts using Eq. \ref{eq: burnslaw}. }
    \label{fig:depol}
\end{figure}

\begin{figure}
    \centering
    \includegraphics[width=\linewidth]{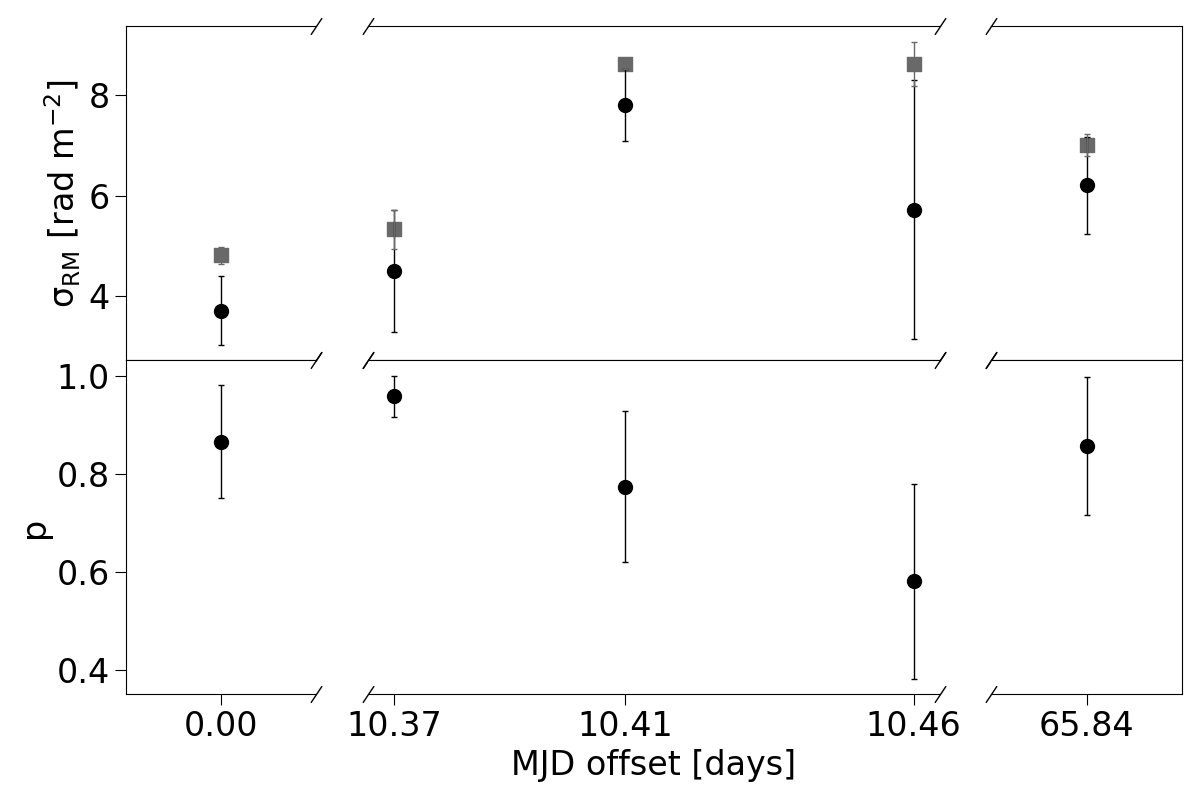}
    \caption{Fitted burns law parameters for selected bursts. `Black circle' points show fitted parameters using equation \ref{eq: burnslaw} and `gray square' points using the un-modified burns law assuming $p$ is unity. Horizontal axis is the MJD offset from the first burst A1.}
    \label{fig:time_depol}
\end{figure}

As shown in Fig. \ref{fig:depol}, the five bursts investigated all show a frequency dependent polarisation fraction, with lower polarisation at lower frequencies. They are all acceptably fit by the Burns depolarisation model, although the lever arm in frequency is not strong in several cases. Crucially, they show significant variation in the degree of depolarisation over timescales $\lesssim$10 days as shown in Fig. \ref{fig:time_depol}.

Given this detection of apparent variability in $\mathrm{\sigma_{RM}}$, we searched for contemporaneous variability in both $\mathrm{|RM|}$ and $\tau$.
We do not see any appreciable change in $|\mathrm{RM}|$ between bursts A1B, M6B, M9 and M24. Furthermore, we cannot constrain a scattering timescale for bursts M24 and P2 so are unable to test if scattering variability is correlated with variability in $\mathrm{\sigma_{RM}}$.  

As discussed in Section \ref{sec:scatter_var}, significant variability in scattering times is observed on timescales of minutes to hours. Therefore, we investigated any potential variations of $\mathrm{\sigma_{RM}}$ within this sub-sample of MeerKAT bursts. To constrain $\mathrm{\sigma_{RM}}$ we assumed that each burst possessed an intrinsic polarisation fraction of $p$ = 1.0 (results are shown in Fig. \ref{fig:meerkat_p1.0_depol_mosaic}). The correlation between the scattering time and $\mathrm{\sigma_{RM}}$, illustrated in Fig. \ref{fig:meerkat_tau_feng}, is weak with a Spearman coefficient of $\rho$ = 0.50 and a p-value of 0.14. Thus, we cannot confidently claim that these two effects are associated with the same environment.

\subsection{Burst separation}
\label{sec:burst_separation}

\begin{figure}
    \centering
    \includegraphics[width=\linewidth]{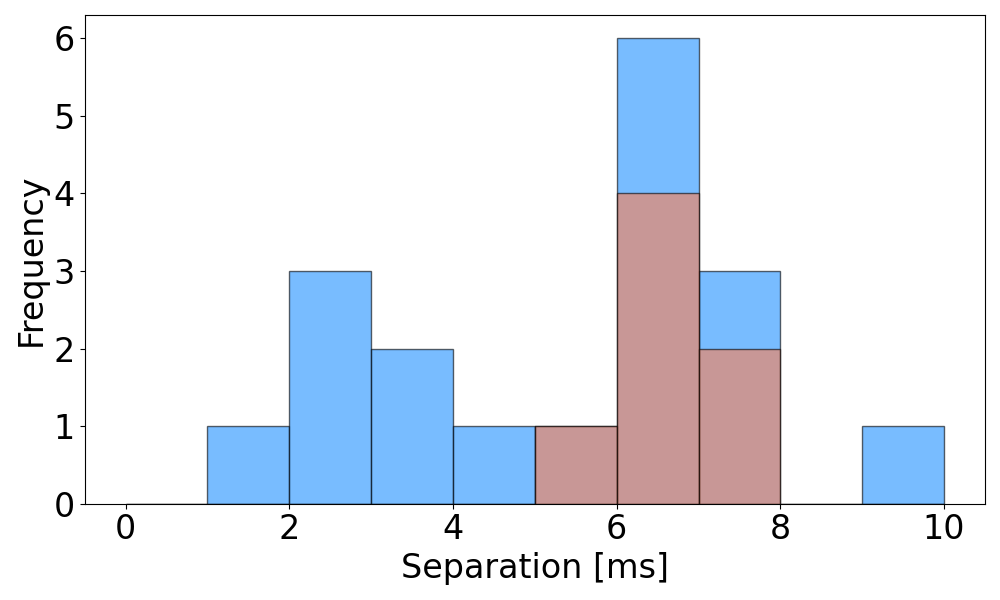}
    \caption{Distribution of component separation for multi-component burst sample. The `blue' histogram shows the distribution of the full sample whereas the `red' histogram only includes bursts with two components.}
    \label{fig:period_hist}
\end{figure}

In Fig. \ref{fig:dynspec_multicomp_mosaic}, we show the sub-sample of bursts with multiple distinct components, which include bursts M6, M8, M15, M24, A1, A2, P5, P7, P8 and P10. Qualitatively, the separation between individual burst components are very similar across the sample. To quantify this we isolated the sub-components of each burst and measured their centroids, which we define as the timestamp at the midpoint of the component's extent; this was performed as an alternative to modelling each component, as some bursts (such as M24) would have required an extremely large number of components, while others (such as M15) had very low S/N for one component. The separation between adjacent components was then measured, and the distribution of component separations is shown in Fig. \ref{fig:period_hist}. The separation between components across all ten multi-component bursts spans almost an order of magnitude, but if we restrict the sample to the 7 bursts that show exactly two components, an extremely narrow separation distribution of 6.8$\pm$0.8 ms is seen. It is noteworthy that no components were observed to be separated by more than 8 ms or less than 5 ms -- i.e., two-component bursts exhibit a strongly preferred component separation of $\sim$7 ms. Other repeating FRBs show similar millisecond timescale separations between sub-components such as FRB20121102A \citep{zhang2018fast}, FRB20190520B \citep{zhang2025statistical}, FRB20201124A \citep{zhou2022fast, zhang2022fast}, FRB20220912A \citep{zhang2023fast} and FRB20240114A \citep{zhang2025magnetar}. However, these distributions span a much larger range in separation. Exploring potential emission mechanisms that prefer such a strict sub-component separation is beyond the scope of this study, but may be related to quasi-periodic microstructure in NS emission \citep{kramer2024quasi} and its relation to the spin period of the source. If the component separation was related to quasi-periodicity in NS emission, it would infer a spin period of $\sim 7$\,s, based on the relationship established in  \citet{kramer2024quasi}.

\subsection{Unique PA structures}
\label{sec:unique_pa}

Bursts A2 and M24 display unique features in their PA profiles, as shown in Fig. \ref{fig:ASKAP_pol_mosaic}. Firstly, A2 exhibits a strong increase in PA on the leading edge of the burst before smoothly declining by $\sim$ 60-70 deg across the first sub-component. The burst also shows clear pulse broadening with a scattering timescale of $\tau$ = 0.020 $\pm$ 0.002 ms. It is possible that the smooth PA decline is a result of diffractive scattering smoothing across an intrinsic PA swing \citep{karastergiou2009complex}. 

Panel (c) of Fig. \ref{fig:ASKAP_pol_mosaic} shows a notable PA jump of $\sim$100 deg in the leading bright component of burst M24. Additionally, during the PA jump the total polarisation falls to $\sim$0 per cent. These are characteristics typically seen in incoherently superimposed orthogonal polarisation modes \citep{manchester1975observations, oswald2023pulsar1} that have been attributed to birefringence in magnetospheric plasma \citep{barnard1986wave}. Collectively, the temporal-polarimetric features of these bursts support a magnetospheric origin for FRB\,20250613A, as posited for several other repeating FRB sources.

\section{Discussion}
\label{sec:DISCUSSION}

The apparent variability in scattering times reported in this work have been observed in several other repeating FRBs. For instance, FRB 20190520B \citep{ocker2023scattering} and FRB 20240114A \citep{panda2025low} exhibit apparent changes in scattering over timescales of minutes to hours, while CHIME and LOFAR studies of FRB 20180916B \citep{sand2023chime, gopinath2024propagation} reveal significant scattering variations on timescales of days. Generically, this scattering variability has been ascribed to a turbulent and ``clumpy'' environment local to the FRB source on AU to sub-AU spatial scales \citep{ocker2023scattering}. This picture is further supported by the large, long-term RM variations observed in repeating sources like FRB 20190520B \citep{anna2023magnetic} and FRB 20201124A \citep{xu2022fast} that vary on timescales ranging from months to years and which imply a changing sightline through a highly magnetised circum-burst medium.

\subsection{NS-Be star binary model}
\label{sec:BEstar}

Analogous systems with changing sightlines to a compact object though a turbulent, highly magnetised medium can be found in the Milky Way. PSR\,B1259$-$63 is a well studied binary pulsar with a companion main-sequence Be star (SS 2883) \citep{johnston1996radio}. The binary system has a highly eccentric orbit ($e\sim0.9$) with a period of 3.5 years. During the periastron passage many of the system's properties undergo drastic changes. The RM increases in magnitude from $\sim$+20 rad m$^{-2}$ to $<-6600$ rad m$^{-2}$, the DM increases by $\sim$10 pc cm$^{-3}$ \citep{johnston1996radio}, and most notably, the scattering timescale changes by a factor of $\sim$20-50 over a month \citep{johnston20011997}. Variations in scattering of up to a factor of $\sim$10-20 are also observed on timescales $<$1 day. Models that incorporate one or both of a stellar wind or geometrically thin and very dense circumstellar decretion disk, with a stellar surface electron density of $\mathrm{n_{e,0}}\gtrsim$10$^{8}$ cm$^{-3}$ \citep{melatos1995stellar}, can broadly reproduce the observed  changes in  propagation effects during the periastron passage. 

More recently, the repeating source FRB\,20201124A has also been suggested to originate from a magnetar in a binary orbit with a Be star companion. FAST observations of the source have shown the RM undergoes a rapid change over $\sim$40 days, beginning with an increase of $\sim$ 200 rad m$^{-2}$ followed by a sharp decrease of $\sim$ 500 rad m$^{-2}$ before returning to a nominal RM of $\sim$ 600 rad m$^{-2}$ \citep{wang2022repeating}. The RM evolution is proposed to originate from a highly eccentric ($e\sim$0.75), 80 day orbit, passing through the disk. Further monitoring of RM variations also suggest the stellar disk precesses \citep{shan2025disk}.

We now investigate whether the observed properties of FRB 20250613A can be self-consistently reproduced by a similar model, namely a progenitor embedded in the clumpy stellar wind of a Be star companion. Consider a NS placed at an offset of 0.5 AU from a Be star with a typical mass of $M_{*}$ = 8 $M_{\odot}$ and radius of $R_{*}$ = 5 $R_{\odot}$. The stellar wind is made up of three primary components: a dense and geometrically thin decretion disk, a smooth axisymmetric stellar outflow, and a random distribution of clumps embedded in the axisymmetric outflow. Initially, we assume that the sightline to the source does not pass through the disk, and hence we ignore this component. The axisymmetric stellar outflow has the following density profile

\begin{equation}
\label{eq:density_model}
    \mathrm{n_{e}}(d_{Be}) = \mathrm{n_{e,0}}\bigg(\frac{R_{*}}{d_{\mathrm{Be}}}\bigg)^{2},
\end{equation}

where $d_{\mathrm{Be}}$ is the distance from the Be star and $n_{e,0}$ is the stellar surface density. We use a typical surface density of $\mathrm{n_{e,0}}$ = 1$\times$10$^{8}$ cm$^{-3}$ \citep{melatos1995stellar}. As the NS orbits through the stellar wind, the strong magnetic field drives a NS wind which produces an electron cavity within a distance of $d_{\mathrm{NS,w}}$ from the NS. We will ignore interactions between the termination shock and the stellar wind and assume the wind extends out to $d_\mathrm{{NS,w}}$ = 0.1 AU \citep{takata2009probing}. We define $\theta$ as the angle between the LOS and the Be star. A schematic of this scenario is depicted in Fig. \ref{fig:nonlinear_diagram}. The DM contribution from this smooth component of the stellar wind can be calculated by integrating along the LOS:

\begin{equation}
\label{eq:DM_s}
    \mathrm{DM}_{s} = \mathrm{n_{e,0}}R_{*}^{2}\int^{L}_{d_{NS}}{\frac{1}{r^{2} + D^{2} - 2Dr\cos{\theta}}}dr,
\end{equation}

where $D$ = 0.5 AU is the binary separation, $r$ the distance from the NS and $L$ is the path length. Here we assume $L \gg d_{NS,w}$ (we set $L$ to 1000 AU during the numerical integration). Starting with a uniform stellar wind outflow, we obtain a stellar wind DM contribution of DM$_{s}$ = 0.44 pc cm$^{-3}$ for a LOS away from the Be star ($\theta = \pi$), and a DM$_{s}$ of 1.63 pc cm$^{-3}$ with a LOS passing closer to the Be star ($\theta = \pi/4$). 

Observations of similar binary systems have demonstrated the size of clumps range from $\sim$10$^{10}$ - 10$^{11}$ cm \citep{moffat1994clumping} within 1 AU of the stellar surface. The clumps occupy a fractional stellar wind volume of $\sim f$ with electron densities around $f^{-1}$ times the average background stellar wind which is referred to as the filling factor. Accounting for the contribution of the clumps on top of the smooth stellar wind, the total stellar wind DM contribution, DM$_{w}$, is DM$_{w} = (2-f)$DM$_{s}$. We can estimate $f^{-1}$ using the large rapid variation in scattering time, which as shown in Section \ref{sec:scatter_var} for the MeerKAT sample, changes by at least a factor of 50 on minute to hour timescales. Using Eq 3 from \cite{johnston20011997}, i.e. $\mathrm{\delta n_{e}^2} \propto \tau^{5/6}$ and a modulation index of $\mathrm{\delta n_{e}^2}/\big<n_{e}\big>^{2} \sim$1 (see \cite{johnston20011997} and references therein), then the electron density in the stellar wind must be changing by at least a factor 5.1. Hence, a filling factor of $f^{-1}\sim$ 5.1, which fits within the level of clumpiness observed for other Be star systems  \citep{sundqvist20182d}, which find  1$\lesssim f\lesssim$10; we note that our estimate of $f^{-1}$ is a lower limit due to upper limits on $\tau$ for narrow bursts. Accounting for the average contribution of clumps, the mean value of DM$_{w}$ is between 0.8-3.0 pc cm$^{-3}$ for a range of $\theta$ values between $\pi/4$ and $\pi$. The actual value observed for any given burst would depend on the number of clumps encountered.

This DM$_{w}$ component is thus the variable component of the total burst DM, with the remainder (``DM$_{\mathrm{ext}}$") originating in the host galaxy ISM and halo, the IGM, and the Milky Way ISM and halo, and remaining constant for all bursts. We cannot measure DM$_{\mathrm{ext}}$ precisely, but can estimate it based on the lower edge of the observed DM distribution. Assuming DM$_{\mathrm{ext}}$ = 173 pc cm$^{-3}$ based on the lower edge of the observed DM distribution yields a range of inferred DM$_{w}$ across the whole MeerKAT sample of 0.7-5.2 pc cm$^{-3}$. This is broadly consistent with the range of values for $\mathrm{DM}_w$ calculated for D = 0.5 AU and a range of viewing angles, although a somewhat higher value of $n_{e,0}$ would be favoured for many combinations of D and $\theta$.

\begin{figure}
    \centering
    \includegraphics[width=\linewidth]{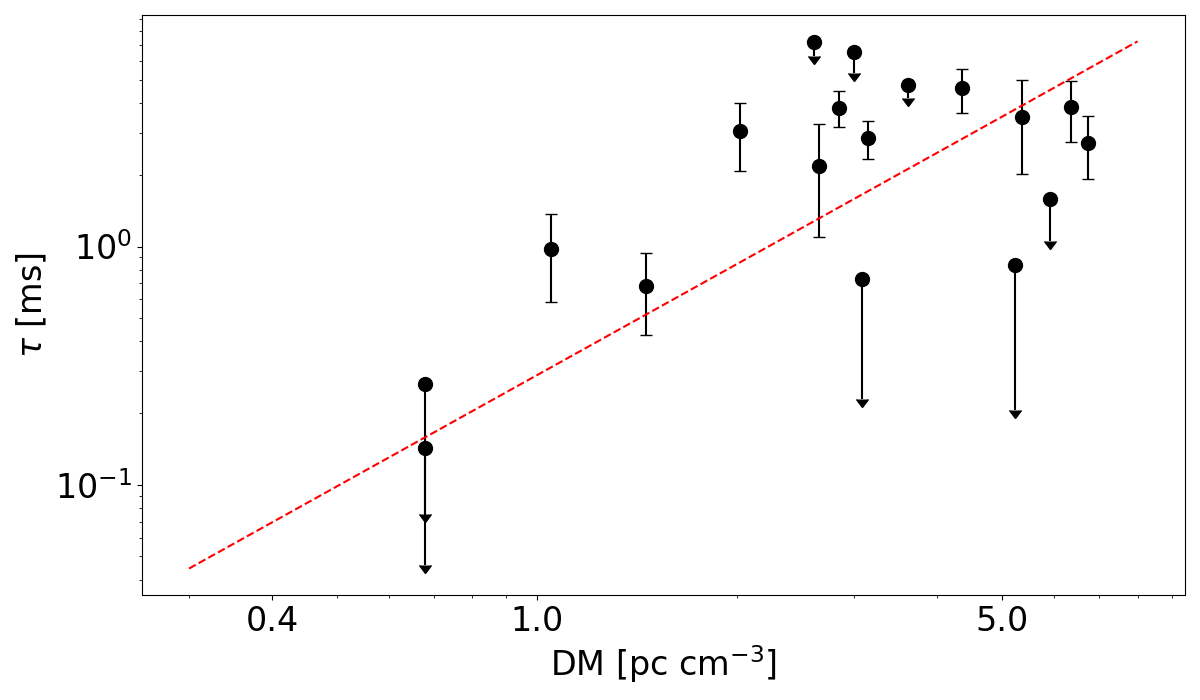}
    \caption{Power law fit for $\tau\propto$ DM$^{\beta}$ for the 24 June MeerKAT burst sample. The red dashed line shows a best fit of $\beta$ = 1.8 $\pm$ 0.2 assuming an external DM of DM$_{\mathrm{ext}}$ = 173 pc cm$^{-3}$. }
    \label{fig:tau_vs_DM}
\end{figure}

For multipath propagation we expect the scattering timescale to correlate with the dispersion measure from the circum-stellar medium DM$_{w}$, i.e. $\tau \propto \mathrm{DM^{\beta}}$ where $\beta$ = 12/5 \citep{johnston20011997}. A similar relationship can be seen in Fig. \ref{fig:tau_vs_DM} albeit with a slightly shallower slope of 1.8$\pm$0.2. The derived slope is extremely sensitive to the DM$_{w}$ at $\tau = 0$, which is poorly constrained, since additional scattering could still occur in a second screen further from the FRB progenitor. The scatter around the mean relation natutally results from the clumpiness in the stellar wind, as each burst will encounter a different number of clumps and lead to a different value of DM$_{w}$. Clumps encountered closest to the stellar companion have the greatest contribution to DM; the DM contribution for a clump at a distance $r$ from the Be star is:

\begin{equation}
    \mathrm{DM_{c}} = 2.6\times10^{-11}\mathrm{n_{e,0}}f^{-1}\bigg(\frac{r}{\mathrm{AU}}\bigg)^{-2} \hspace{0.1cm} \mathrm{pc\hspace{0.1cm}cm^{-3}}.
\end{equation}

For $r$ = 0.5 AU, DM$_{\mathrm{c}}$ = 0.053 pc cm$^{-3}$. For a LOS that passes closer to the Be star, $r$ = 0.3 AU. then DM$_{\mathrm{c}} \sim$ 0.13 pc cm$^{-3}$. These variations in scattering time and DM should occur on timescales of order the transit time of a clump crossing the LOS. For a clump size of 0.01 AU and a typical stellar wind velocity of $v_{\mathrm{w}}\sim$1000 km/s \citep{beskrovnaya2025stellar} the variations would be occurring over timescales of $\sim$1500 seconds which is close to the average wait time of bursts in the MeerKAT sample ($\sim$1100) seconds.

Given the high electron density encountered in sightlines that pass close to the Be star, low radio frequencies may be attenuated due to free-free absorption. The optical depth for free-free absorption is given by:

\begin{equation}
    \tau_{\mathrm{ff}} = 4.0\times10^{-7}\bigg(\frac{\nu}{\mathrm{GHz}}\bigg)^{-2.1}\bigg(\frac{T_{e}}{\mathrm{K}}\bigg)^{-1.35}\int^{L}_{0.1}{n_{e}(d_{\mathrm{Be}})^{2}}dl,
\end{equation}

where $\nu$ is the burst frequency and $L$ the path length in AU away from the NS (which we set to 1000 AU as was the case for equation \ref{eq:DM_s}). Including the stellar wind clumps the optical depth is $\tau_{\mathrm{ff}}^{'} = (f^{-1}+1-f)\tau_{\mathrm{ff}}$. Using a burst frequency of $\nu$ = 1.0 GHz, a binary separation of $D$ = 0.5 AU and a lower-limit stellar wind photo-ionised temperature of $T_{e}\sim$10$^{4}$ K \citep{snow1982stellar}, then $\tau_{\mathrm{ff}}$ ranges from 0.04 to 0.8 for $\theta$ between $\pi$ and $\pi/4$. Given that this entire range of sightlines remains optically thin, and that the temperature in the wind is likely to be higher than the fiducial value used above, we conclude that in most cases the stellar wind will not lead to substantial free-free absorption. Sightlines that pass extremely close to the Be star, or that pass through the stellar disk, could however become optically thick due to free-free absorption.

The observed RM evolution of FRB\,20250613A offers another opportunity to test for consistency with the Be star companion model. Over the short term, we would expect variations in the clumpy sellar wind to induce short-term stochastic variations in RM (due to the changing density along the sightline), whilst the orbital motion of the binary system will cause  larger RM variations over longer periods as the magnetic field strength and orientation along the sightline changes. On both short and long timescales, RM evolution consistent with these expectations is seen.

On minute to hour timescales, the source exhibits RM changes of order tens of rad m$^{-2}$.
The RM contribution for a clump of electrons at a distance r from the stellar surface is:

\begin{equation}
    \mathrm{RM} \simeq 8.2\times10^{5}\bigg(\frac{B_{0}}{\mathrm{G}}\bigg)\bigg(\frac{r}{\mathrm{R_{*}}}\bigg)^{-2}\bigg(\frac{\mathrm{DM}}{\mathrm{pc\hspace{0.1cm}cm^{-3}}}\bigg),
\end{equation}

If we assume a radial magnetic field topology (neglecting any dipolar and/or toroidal components) with a surface magnetic field strength of B$_{0}$ = 10 G \citep{wang2022repeating}, take r = 0.5 AU, and assume a DM contribution of $\sim$0.024 pc cm$^{-3}$ based on a clump with a transverse size of 0.01 AU and an electron density of $f^{-1}\mathrm{n_{e}}$, where $\mathrm{n_{e}} \sim 10^5$ at a distance r, then we obtain stellar clump RM contributions of $\sim400$ rad m$^{-2}$. Given that the observed RM variations on short timescales are an order of magnitude smaller, the surface magnetic field should be somewhat lower than 10 G. This is consistent with limits on the magnetic field strength of Galactic Be stars \citep{Wade2016}.

\begin{figure}
    \centering
    \includegraphics[width=\linewidth]{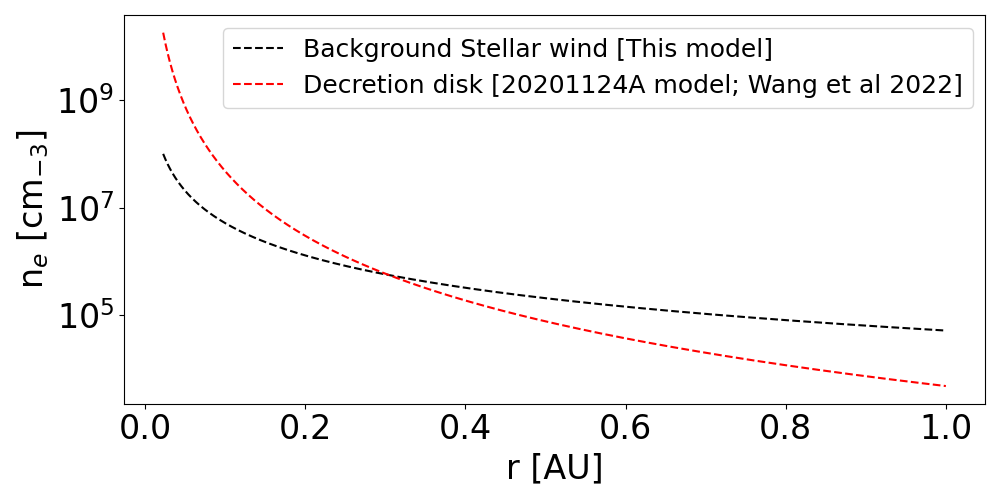}
    \caption{Stellar wind density profile. The black line shows the radial electron density profile of an out-flowing stellar wind shown assuming a stellar surface density of $\mathrm{n_{e,0}} = 1\times$10$^{8}$ cm$^{-3}$. The red line is the on-plane density profile of the stellar disk used in \protect\cite{wang2022repeating} which has a surface density of $\mathrm{n_{e,0}} = 1.8\times$10$^{10}$ cm$^{-3}$ and a power-law index of $\beta$ = 4.}
    \label{fig:SW_density}
\end{figure}

We see more significant variation of $\Delta$|RM|$>$200 rad m$^{-2}$ on monthly timescales. Similar variations are seen in the repeating FRB\,20201124A and are consistent with a changing sight line through an ordered magnetic field. In general, Be stars possess both an out-flowing stellar wind and a compact disk. However, the mass profile in the disk can differ markedly between different sources. A typical disk has a surface electron density of $\sim$10$^{11}$ - 10$^{14}$ cm$^{-3}$ with a radial power-law $\beta$ of $\sim$2-4 \citep{rivinius2013classical}. The RM profile in \cite{wang2022repeating} was best fit with a relatively weak but steep stellar disk which is shown in Fig. \ref{fig:SW_density}. When compared to the stellar wind model we have used until now, the disk starts off with highest density at the stellar surface, which quickly drops below the background stellar wind at $\sim$0.3 AU. This means for our model of FRB\,20250613A we only see the impact of the disk over the full orbit which will result in the long timescales changes we see in RM as supported by \cite{johnston1996radio} and \cite{wang2022repeating}. Consequently, we ignore the disk for our calculations later.

Given that the mean |RM| is over an order of magnitude larger than the RM variations, the sight-line cannot be varying drastically with respect to the mean B field orientation over this time, at least in the region of plasma that dominates the total RM. This implies that either the orbital period is considerably longer than the observation time, or the majority of the observed RM originates in another dense ionised region along the line of sight (e.g., in a surrounding supernova remnant). Our current burst sample does not allow us to constrain the putative binary period, other than a likely lower limit of $\gtrsim$100 days 
given the timespan of our observations.
We note that binary motion with a comparable period to this lower limit has been suggested as a cause of the activity windows seen in FRB\,20121102A \citep{rajwade2020possible}.

Finally, we can consider whether the spectral depolarisation observed in FRB\,20250613A is consistent with this model. Spectral depolarisation arises from multipath-propagation through turbulent environments and is thus expected to for radiation propagating through a Be star wind \citep{johnston1996radio}. Following \cite{wang2022repeating}, we assume that $\mathrm{\sigma_{RM}}$ arises from variations in electron density which lets $\mathrm{\sigma_{RM}}=f\mathrm{|\Delta RM|}$. Stochastic variations in Faraday rotation result in spectral depolarisation when |$\Delta$RM| > 1/($f\mathrm{\lambda^{2}}$). Assuming $\Delta$RM = 200 rad m$^{-2}$ and a central frequency of 1.0 GHz then the filling factor must be $<$18 or the fractional volume of the clumps $>$ 0.055 for depolarisation to come from the stellar wind. Since we derive a lower limit of $f^{-1} >$ 5.1 we conclude the spectral depolarisation observed in FRB20250613A is consistent with originating in a stellar wind. FRB20250613A also shows temporal changes in $\mathrm{\sigma_{RM}}$ over timescales of tens of days, consistent with a changing line of sight that intersects a changing ensemble of wind clumps. Consequently, we should also expect to observe a change in $\mathrm{\sigma_{RM}}$ over timescales $<$1 day as clumps closer to the source will have a much greater influence in depolarising the bursts. Such variations are allowed, but not demanded, by the current sample of MeerKAT bursts (see Fig.~\ref{fig:depol}).

We note that our proposed progenitor model for FRB\,20250613A is also consistent with expectations from the global environment of its host galaxy. There is a comparatively larger known population of Be X-ray binaries (BeXRBs) in the sub-solar metallicity Small Magellanic Cloud (SMC) than in the higher-metallicity environments of the Large Magellanic Cloud (LMC) or Milky Way \citep{coe2015catalogue,HaberSturm2016}, which has generally been attributed to a metallicity dependence on the X-ray luminosity function for both BeXRBs and HMXBs more generally \citep{Douna2015,Lehmer2021,Liu2024}. While the strength of this metallicity dependence is unclear, the fact that FRB20250613 occurs in a sub-solar metallicity host galaxy is self-consistent with BeXRB populations in nearby sub-solar metallicity galaxies, and with the Be-NS star binary interpretation.

\subsection{Non-linear propagation effects}
\label{sec:nonlinear}

The previous section showed that qualitatively, all the temporal variability seen on timescales of minutes to months can be explained in the Be star model. However, a handful of bursts from this source show startling variations in the inferred scattering time and dispersion measure on a timescale of milliseconds, which cannot be explained by a changing line of sight through a putative companion star's wind. As shown on the left panel of Fig. \ref{fig:ASKAP_pol_mosaic}a, the leading component of burst A1 is best fit with a 30-fold greater value of scattering, a 0.27 pc cm$^{-3}$ higher dispersion measure and is significantly more depolarised. Additionally, the first component has a rotation measure $\sim-$50 rad m$^{-2}$ larger in magnitude compared to the second component of the burst (see Table \ref{tab:burst_tab2}). For these variations to occur as a result of a changing sightline through a clumpy medium requires implausibly small clumps, or the source motion to be highly relativistic. Taking the same 0.01 AU clump size used in the previous section (which corresponds to $\sim$5 light seconds), having two successive bursts separated by 7 milliseconds probe entirely different sightlines requires the emitting object to have an apparent transverse motion of $\sim$100$c$, where $c$ is the speed of light. Such apparent superluminal motion would require highly relativistic motion of the emission region along a direction close to the line of sight, sustained over a length scale of (at a minimum) many astronomical units; a physical picture that is in tension with the magnetospheric origin supported by other aspects of the FRB\, 20250613A emission as already discussed.

However, a changing sightline is not the only way in which the propagation effects can vary from component to component, and in fact a relevant mechanism is {\em expected} to operate at sub-AU distances between the progenitor and scattering screen postulated in the Be star companion model. Up to a distance of $\sim$AU from the origin of the FRB emission, the average FRB luminosity is large enough to excite the free electrons in the plasma to relativistic speeds via a ponderomotive force that drives nonlinear propagation effects \citep{yang2020fast, zhang2023physics, lyutikov2024escape}. This occurs when the strength parameter (also known as the wave-breaking amplitude \citep{katsouleas1988wave}) is: 

\begin{equation}
\begin{split}
a_{0} & = \frac{eE_{0}}{m_{e}c\omega} \\ & = \sqrt{\mathrm{I}[\mathrm{W cm^{-2}}]\hspace{0.1cm}\lambda^{2}[\mathrm{\mu m^{2}}]/1.38\times10^{18}[\mathrm{Wcm^{-2}\mu m^{2}}]} \geq 1.0,
\end{split}
\end{equation}

where $\mathrm{I} = L_{\mathrm{iso}} / 4\pi r^{2}$ is the intensity of the burst at a distance $r$ from the source \citep{king2023perspectives}. The acceleration time is a fraction of the total pulse width and so the bulk of the pulse will not interact with the electrons in a cold (non-relativistic) state. For electrons that have been accelerated to a lorentz factor of $\gamma$ where $\gamma \simeq \sqrt{1 + a_{0}^{2}/2}$, their relative mass increase leads to a reduction in the plasma frequency of 1/$\gamma$, reducing the effects of Faraday rotation and dispersion by $\sim\gamma$, and reducing scattering by a factor of $\sim\gamma^2$. Beyond some distance $d$ away from the source, $a_0$ falls to a value less than 1. In this regime the ``quiver" velocity of the electrons is much smaller than $c$. As a result they undergo simple harmonic motion where the energy loss is relatively small, and the electrons are no longer accelerated to relativistic speeds in the direction of the FRB pulse. Since $\gamma \sim$ 1 in this regime, the propagation effects are comparable to a cold plasma. The subsequent discussion considers only the acceleration of the electrons to relativistic speeds; other possible manifestations of the non-linear propagation effects such as self-focusing and filamentation \citep{sobacchi2022filamentation, zhang2023physics} are not considered for simplicity.

We consider the effect that these nonlinear effects should have on the propagation of successive pulse components to qualitatively explain the short ($\sim$10 millisecond) timescale variation in a number of propagation effects such as scattering time and DM that we see in the morphology of multi-component bursts such as A1. Consider a two component burst with luminosities $L_{1}$ and $L_{2}$, a common width $w$ and a separation $\Delta t$ propagating in the stellar wind of the Be star (a diagram of the scenario is shown in Fig.~\ref{fig:nonlinear_diagram}). Initially, the strength parameter of the first component $a_{0,1}$ > 1 and so the electrons are efficiently accelerated to relativistic speeds, and the pulse experiences reduced dispersion and scattering. Energy is expended by the pulse accelerating the electrons in the plasm, and so $a_{0,1}$ drops more rapidly than $1/r^2$. At some point $d_{1}$,  $a_{0,1}$ has reached unity, and ``normal" cold plasma propagation effects resume for the first pulse component for the remainder of its journey through the stellar wind.

After the passage of the first pulse component, the remaining electrons encountered by the second pulse component ($\Delta t$ seconds later) are already traveling at relativistic speeds. Thus, even if the second pulse component had an identical initial luminosity to the first component, it does not lose as much energy accelerating the plasma with its passage, and by the time it reaches $d_{1}$, the strength parameter of the second component $a_{0,2}$ would still be greater than unity. It can then continue to propagate to a second distance $d_2$ before $a_{0,2}$ reaches unity and the reduction in dispersion and scattering from electron acceleration ceases. Any wind clumps encountered between $d_{1}$ and $d_{2}$ would thus impact less additional dispersion (and significantly less additional scattering) on the second pulse component, compared to the first. Beyond $d_{2}$, both pulse components propagate normally.

\usetikzlibrary{calc, shadings, angles, quotes, arrows.meta}

\newcommand{\laserpulse}[3]{

    \fill[left color = white, right color = white, middle color = red, draw = white] (#1, #2) ellipse (0.75*#3 and 0.0975)
}

\begin{figure*}
    \centering

\begin{tikzpicture}





\node[inner sep=0pt] (a0_plot) at (15.0, 0.0)
    {\includegraphics[width=0.40\textwidth]{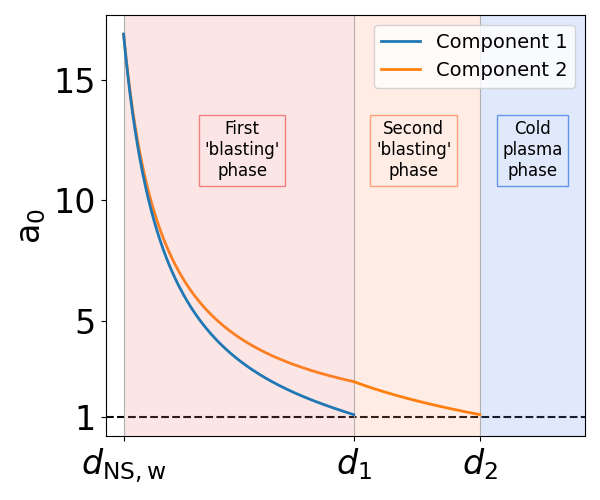}};
\draw[inner color=orange, opacity=0.9, draw=white] (8,0) circle (3cm);

\draw[color=white, line width = 1.0mm] (8.5, 1.5) coordinate (A) -- (10.0, 0.0) coordinate (B);
\draw[color=white, line width = 1.0mm] (8.0, 2.0) coordinate (A) -- (8.5, 1.5) coordinate (B);
\coordinate [label=left:{$\mathrm{d_{1}}$}] (A) at (8.5, 1.5);
\node[below=2.0pt] at (9.0, 0.0) {D};

\begin{scope}
    \clip(8.0, 0)circle(3.0cm);
    \draw[fill=white, rotate=45] (7.05, -8.65) ellipse (1cm and 2cm);
    
\end{scope}

\draw (8.47, 1.47) coordinate (A) -- (9.97, -0.03) coordinate (B) -- (8.0, 0.0)               coordinate (C)
    pic ["$\theta$", draw, angle radius = 1.1cm] {angle};
\draw [dashed, |->] (8.47, 1.47) coordinate (A) -- (7.27, 2.67) coordinate (B);

\draw (8.03, 2.03) coordinate (A) -- (10.03, 0.03) coordinate (B);
\draw [dashed, |->] (8.03, 2.03) coordinate (A) -- (7.53, 2.53) coordinate (B);
\coordinate [label=right:{$\mathrm{d_{2}}$}] (A) at (8.03, 2.03);
\draw[fill=white, draw=white] (8,0) circle (0.4cm);
\draw[fill=cyan, opacity=0.5] (8,0) circle (0.4cm);

\draw[opacity=1.0, fill=white] (10.0,0) circle (0.2cm);



\node[font=\large, below=2.0pt] at (6.0, -2.0) {Stellar Wind};
\node (1) at (6.0, -2.0) {};
\node (2) at (-1.7+8, -1.4) {};
\draw [line width = 0.5mm,->] (1) to[out=80,in=225, looseness=1.5] (2);

\node[font=\large, below=2.0pt] (1) at (-0.6+8, -0.6) {Be Star};
\node (2) at (-0.2+8, -0.2) {};
\draw [line width = 0.5mm,->] (1) to (2);

\node[font=\large, below=2.0pt] (1) at (2.9+8, -0.1) {NS};
\node (2) at (2.1+8, -0.1) {};
\draw [line width = 0.5mm,->] (1) to (2);

\node[font=\large, below=2.0pt] (1) at (2.88+7.8, -1.5) {NS Wind};
\node (2) at (1.88+8, -1.35) {};
\draw [line width = 0.5mm,->] (1) to[out=115, in=45] (2);

\end{tikzpicture}
\caption{Schematic of the stellar wind toy model. The Be star (shown in blue) creates an isotropic stellar wind (shown in orange) with a mass density of 1/$r^{2}$ (depicted by the colour gradient). A NS (shown as a white circle) orbits the Be star with a binary separation of D. The NS magnetic field drives electrons away from the immediate vicinity of the NS creating an electron cavity where the effective electron density is zero. The boundary of this "NS wind", where the magnetic force of the NS balances out the radiative force of the stellar wind is called the termination shock. A 2-component burst is emitted at an angle $\theta$ away from the Be star. For a distance $\mathrm{d_{1}}$ away from the NS the first component blasts through the stellar wind (shown as a thick white line), accelerating the line-of-sight electrons to relativistic speeds and reducing their effects on the propagation of both burst components to a negligible level. We term this section of the line of sight the first blasting phase, and in this phase the intensity of the first component falls away faster than $1/d^2$ as it loses energy accelerating the stellar wind. After a distance $d_{1}$ away from the NS the first component no longer accelerates electrons to relativistic speeds and begins to undergo scattering, dispersion, Faraday rotation, etc as in a cold plasma (shown as the dashed line). However, at a distance $d_{1}$ the second component is still sufficiently powerful to blast through the stellar wind up to a distance $d_{2}$, and so we term the section $d_{1}$ to $d_{2}$ the second blasting phase, where the second component experiences reduced propagation effects while the first has propagated normally. After a distance $d_{2}$ neither burst component is accelerating electrons to relativistic speeds, and we term this the cold plasma phase. }
\label{fig:nonlinear_diagram}
\end{figure*}

\begin{figure}
    \centering
    \includegraphics[width=\linewidth]{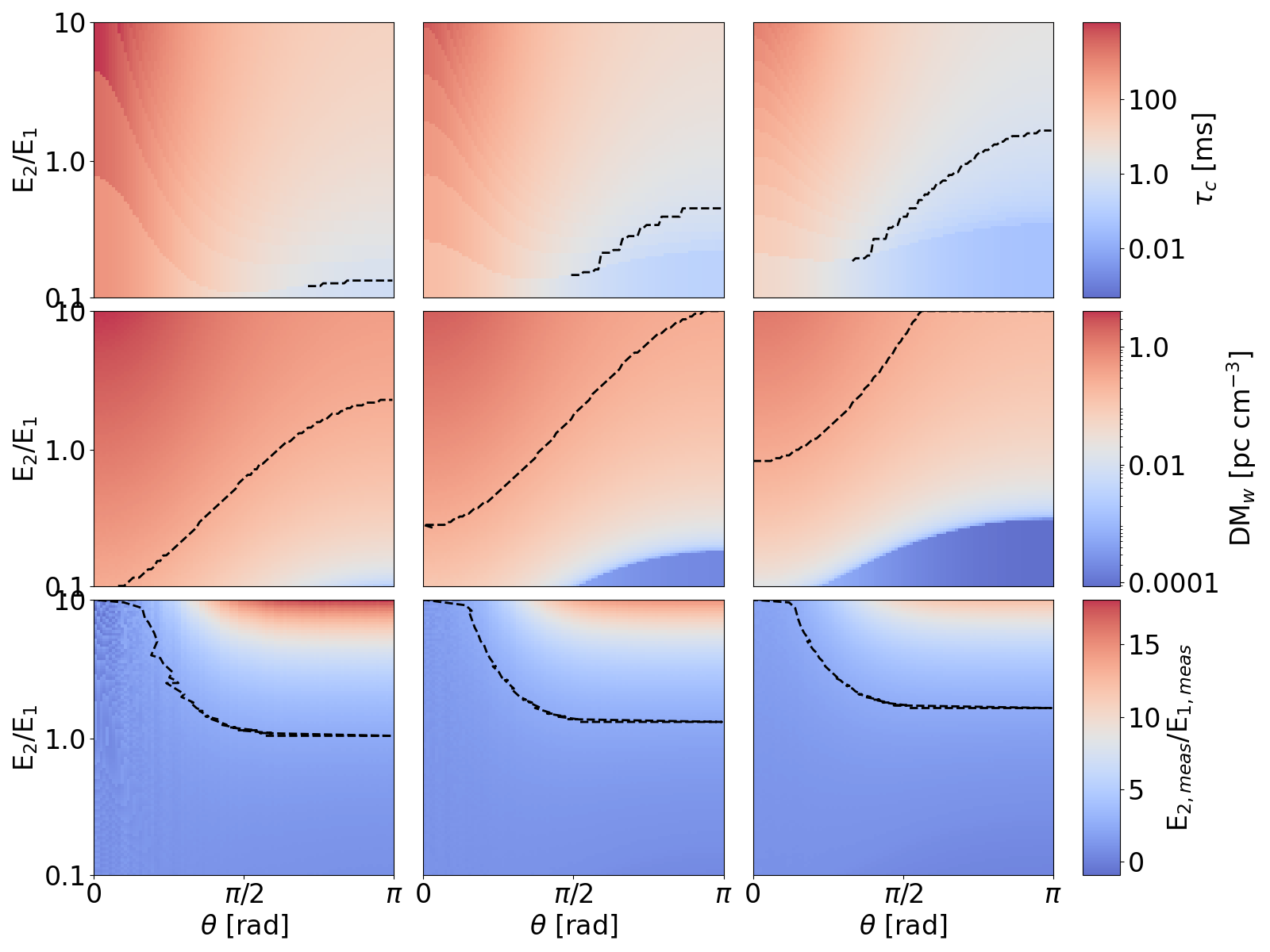}
    \caption{Results of nonlinear simulation. Top panel row: Scattering time induced by clumps in the region $d_{1}$ to $d_{2}$. Black lines shows a scattering time of 1.0 ms. Middle panel row: Integrated DM in the region $d_{1}$ to $d_{2}$. Black lines show a DM of 0.27 pc cm$^{-3}$. Bottom panel row: Observed peak luminosity ratio between the two components at the moment they cease inducing non-linear propagation effects in the stellar wind. Black lines show an observed peak luminosity ratio of 2.62. Panels on the top row show the clump contribution to the scattering timescale and the bottom panels show the combined DM contribution from the clumps and background stellar wind for a range of relative pulse energies and LOS directions. The panel columns from left to right use a binary separation of 0.3 AU, 0.5 AU and 0.8 AU. }
    \label{fig:2p_nonlinear_sim}
\end{figure}

We tested whether the simple model of a Be star stellar wind introduced in Section~\ref{sec:BEstar}, and illustrated in Fig. \ref{fig:nonlinear_diagram}, is consistent with the apparent change in pulse component properties seen in burst A1.
We again assume initially that the stellar wind is spherical and isotropic with a surface density of $\mathrm{n_{e,0}}$ = 1.0$\times$10$^{8}$ cm$^{-3}$, and neglect any dense decretion disk (equivalent to assuming that our line of sight does not pass through the disk). We sample a range of values for the binary separation $D$, the angular offset between the star and the LOS direction $\theta$ and ratio of the initial energy of the two pulse components. We assume an isotropic luminosity $L_{\mathrm{iso}}$ = 10$^{42}$ erg\,s$^{-1}$ for the first pulse component, and assume the sub-components each had an intrinsic width of $w$ = 1 ms. We ignore the intrinsic chromaticity of these effects making all of our calculations at a frequency of 1.0 GHz. We integrate the energy loss along the LOS:

\begin{equation}
\begin{split}
    E_{\mathrm{loss},i} = 4\times 10^{7}\pi \mathrm{n_{e,w}}m_{e}c^2 \delta x r^{2}\sqrt{1 + \frac{E_{i}\lambda^{2}}{8\pi r^{2}k}},
\end{split}
\end{equation}

where $E_{i}$ = $E_{i-1}$ - $E_{\mathrm{loss},i-1}$ and $k$ = $1.38\times10^{18}$ $ \mathrm{W}$cm$^{-2}$ \textmu m$^{2}$; $n_{e,w} = (2-f)n_{e}$ is the mean electron density after accounting for the mean contribution of the stellar wind clumps along with the smooth outflow, and $n_{e}$ is calculated from equation \ref{eq:density_model}. We integrate along a path starting from $d_\mathrm{{NS,w}}$ = 0.1 AU away from the NS to 3 AU with a step size of $\delta$x = 1.67$\times$10$^{-4}$ AU. We integrated the energy loss for the first sub-component until a distance $d_{1}$. Then we integrated the energy loss of the second sub-component until a distance $d_{2}$, at which point the program terminated. We calculated the difference in DM$_{w}$ between the two components by simply integrating the electron density over the segment $d_{2}$ - $d_{1}$. To calculate the difference in scattering time we took the smooth background stellar wind and placed clumps of size 10$^{11}$ cm equidistant along the segment $d_{2}$ - $d_{1}$ with a separation of 0.02 AU (which effectively assumes a filling factor of $f\sim$5.0); this is equivalent to taking a single, idealised realisation of the clumpy stellar wind. We ignored any scattering induced by the background stellar wind since $\tau \propto \mathrm{n_{e}}^{10/6}$ from equation 3 of \cite{johnston20011997}. Fig. \ref{fig:2p_nonlinear_sim} shows the calculated differential DM, differential scattering, and observed peak luminosity ratio as a function of $\theta$ and $E_2/E_{1}$, with the black dashed lines showing the best-fit values in these quantities from burst A1. 

$\Delta$DM is not well constrained for burst A1 as the first component has a DM uncertainty of $\sim$0.7 pc cm$^{-3}$, so there is a large parameter space at a binary separation of $\geq0.5$ AU that can reproduce the variations in scattering (1.00 $\pm$ 0.04 ms) and DM ($0.3\pm0.7$ pc cm$^{-3}$) between the two components. An additional constraint is however provided by the the ratio in the measured peak luminosity between the two components, which we find to be $L_{2}$/$L_{1}$ = 2.62 $\pm$ 0.16 when converting from pulse fluence to luminosity using the inferred intrinsic pulse width (i.e., after de-convolving each component with their respective scattering tail). This relatively high value favours larger separations and LOS geometries that face away from the companion star. As an example, for $\theta$ = 2.7 radians, a binary separation of 0.8 AU and an intrinsic peak luminosity ratio of $L_{2}$/$L_{1}$ = 1.55, the modelled difference in scattering ($\tau$ = 1.03 ms), DM (0.07 pc cm$^{-3}$), and observed peak luminosity ratio ($L_{2}/L_{1}$ = 2.48) are all consistent within 1$\sigma$ to the observed values. In principle, then, this simple model can replicate the observed variations in propagation effects observed in burst A1.

This approach incorporates a number of simplifications whose effects bear examining. Firstly, a small filling factor (i.e., a large fractional volume) is assumed, which means the stellar outflow consists of a large number of clumps (on the order of tens). Independent realisations of the sight-line would see a varying number of clumps, but the fractional variance (and hence the variation in the scattering time for a burst of a fixed luminosity) would be small. However, if a large filling factor is used, the number of clumps along the sight-line will be of order a few and the variations in scattering time could be very substantial across different realisations. Additionally, the clumps expand as they move away from the star, with hydrodynamical simulations showing that the clump-filling factor changes substantially before settling $\sim$2R$_{*}$ from the stellar surface \citep{sundqvist20182d}. The detailed properties of the clumps further from the star are poorly constrained, which translates to additional uncertainty on their impact on the (differential) propagation effects between bursts. Additionally, we do not consider any propagation effects that are induced further away from the source at a distance $d$ > $d_{2}$, although they would impact both burst components identically. For scattering specifically, however, the contribution in this cold plasma regime would be relatively small given the model parameters that  adequately recreate bursts such as A1, while for DM the contribution would be both small and common to both pulse components, and hence indistinguishable from DM$_{\mathrm{ext}}$.

\begin{figure}
    \centering
    \includegraphics[width=\linewidth]{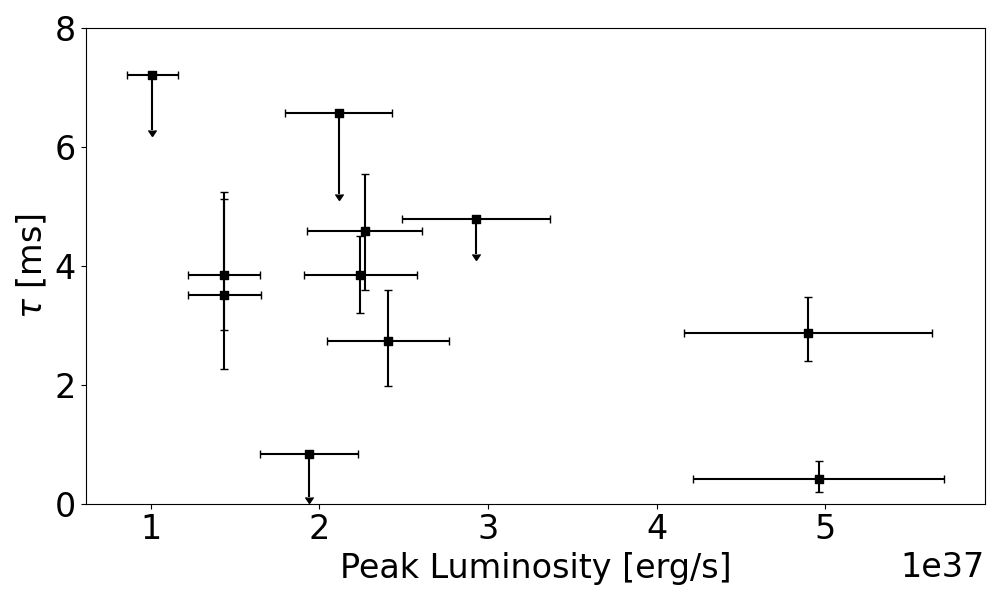}
    \caption{De-convolved peak luminosity vs scattering timescale scaled to 1.0 GHz as described in Section \ref{sec:nonlinear}. Peak luminosity is calculated using the formula $L_{p}$ = $E$/$\sigma \sqrt{2\pi}$ where $\sigma$ is the standard deviation of the un-scattered Gaussian pulse measured by fitting the scattering timescale. MeerKAT bursts with a $S/N$ < 15 have been excluded.}
    \label{fig:energy_vs_tau}
\end{figure}

In this model, the distance over which non-linear effects are relevant is dependent on the intrinsic burst luminosity, and hence a correlation between this and the observed scattering time should be expected, with more luminous bursts experiencing lower scattering. Directly interpreting the observed quantities is challenging, since we  must deconvolve the effects of scattering to obtain the peak luminosity, which is still not the intrinsic luminosity (due to the losses incurred accelerating electrons in the blasting region). These caveats notwithstanding, a negative relationship is still expected between measured burst luminosity and scattering timescale. As demonstrated in Fig. \ref{fig:energy_vs_tau} the MeerKAT burst sample does show a negative -- but not significant -- correlation between the observed burst luminosity and the scattering timescale, with a Spearman coefficient of $\rho =$ -0.455 and a p-value of 0.16.

Additionally, the strength parameter is inversely proportional to the radio frequency, i.e., $a_{0} \propto 1/f$. Lower frequency emission will thus experience greater energy loss.
If the intrinsic burst spectrum is constant between components, we should then expect that the first pulse component is more heavily attenuated at low radio frequencies. In all of the two-component bursts with apparent differential scattering, the first  component is both temporally broader and cut off at higher radio frequency than the second component: see bursts A1,   M6, P7 and P10 in Fig.~\ref{fig:dynspec_multicomp_mosaic}.   

We should also expect this correlation to change as a function of the distance between the burst origin and the location of the peak scattering, which would have both a stochastic component (due to the clumpy wind) and a deterministic component (due to the progenitor orbital phase). The first component of burst A1 has an observed peak luminosity of $L_{p}$ = 1.9$\times$10$^{42}$ ergs/s, which is higher than seen for some MeerKAT bursts, while still exhibiting relatively large scattering compared to those bursts. This would suggest that the electron column traversed through the stellar wind was higher at this time than during the MeerKAT sample. Given the 10 day separation between the ASKAP and MeerKAT observations, this could be the result of orbital motion, or could simply be the result of burst to burst sightline variability in clump sampling.

Qualitatively, the model presented here can explain most of the observed bursts, with the exception of M24 (see Fig. \ref{fig:dynspec_multicomp_mosaic}). This burst displays a complex multi-component morphology as well as unique polarisation and spectral features. In contrast to other multi-component bursts, the first component is the brightest and is fit by relatively low scattering, and subsequent (fainter) components have a larger apparent scattering time, which cannot be explained by non-linear propagation effects (since the plasma should not yet have cooled after the passage of the first burst, meaning the second component should not be more scattered than the first component, despite being fainter). Future studies with a larger sample of bursts from FRB\,20250613A could ascertain whether the morphology displayed by this burst is a rare occurrence plausibly explained by intrinsic burst structure evolving in a similar way to pulse broadening, or is more frequent and poses a challenge to the model presented here.

FRB\,20250613A shares many similarities with FRB\,20201124A, substantial variations in RM on timescales of months, as well as scattering time and DM variations on timescales of minutes to hours \citep{lanman2022sudden}. The energy distribution of FRB20201124A is also extremely broad \citep{zhang2022fast} and allows for bursts with high energies that are necessary for non-linear propagation effects. Therefore, if these two sources share a common progenitor type, we should expect to see burst luminosity dependent propagation effect changes in FRB\,20201124A. However, no bursts matching the archetypical two-component burst from FRB\,20250613A (faint and broad first component, bright and narrow second component) are visible in the published data for this source. FRB\, 20201124A exhibits generally more temporal substructure then we see for FRB\,20250613A, which may complicate efforts to isolate time-dependent propagation effects from intrinsic burst substructure evolution.

We conclude this section by noting that while the observational characteristics of FRB\,20250613A are consistent thus with a Be-NS binary origin, generically the time-varying propagation effects that are observed could be generated by any physical scenario placing a relatively thin, dense, highly magnetised, and clumpy scattering screen at a distance $\lesssim$1 AU from the FRB progenitor.


\section{Conclusion}
\label{sec:CONCLUSION}

We have presented observations of FRB\,20250613A, a repeating FRB discovered by the CRAFT survey on the telescope. The FRB has been localised to a low-mass, low-metallicity star-forming galaxy that is reminiscent of the host of the first repeating FRB, FRB~20121102A. Our monitoring campaign revealed numerous phenomena that provide insight into the progenitor and its surroundings, including substantial variations in scattering on the timescale of minutes, changes in spectral depolarisation on timescales of days, and long term variability of the RM over months. While other explanations are possible, all these properties are consistent with a progenitor embedded in a strong and clumpy wind of a massive star. If this is the case, some bursts from the FRB are sufficiently energetic to accelerate the electrons in the stellar wind to relativistic speeds, inducing non-linear propagation effects that can explain the observed apparent changes in scattering over millisecond timescales.

Longer and more sensitive monitoring of this source with high time resolution and full polarimetry would enable constraints on the geometry of the putative binary orbit and the LOS, potentially confirming (or rejecting) this interpretation of the burst environment.

\section*{Acknowledgements}

We thank Stella Ocker for useful discussion on multipath propagation and pulse broadening functions. MG is supported by the UK STFC Grant ST/Y001117/1. MG acknowledges support from the Inter-University Institute for Data Intensive Astronomy (IDIA). IDIA is a partnership of the University of Cape Town, the University of Pretoria and the University of the Western Cape. For the purpose of open access, the author has applied a Creative Commons Attribution (CC BY) licence to any Author Accepted Manuscript version arising from this submission.
A.C.G. and W.F. acknowledge support from the National Science Foundation grant AST-2206494 as members of the Fast and Fortunate for FRB Follow-up team. W.F. acknowledges support from a CAREER grant No. AST-2047919, the David and Lucile Packard Foundation and the Research Corporation for Science Advancement through Cottrell Scholar Award \#28284.
ATD, RMS and JJ-S acknowledge support through Australian Research Council Discovery Project DP220102305. Murriyang, CSIRO’s Parkes radio telescope, is part of the Australia Telescope National Facility (https://ror.org/05qajvd42) which is funded by the Australian Government for operation as a National Facility managed by CSIRO. We acknowledge the Wiradjuri people as the Traditional Owners of the Observatory site.

\section*{Data Availability}
The data used in this work will be made available upon reasonable request.

\section*{Code}
The majority of the data analysis and plotting in this work used the ILEX python package for FRB analysis, which can be accessed at https://github.com/tdial2000/ILEX.



\bibliographystyle{mnras}
\bibliography{ref} 




\appendix

\subsection{Appendix: A - Burst data tables}

\renewcommand{\thetable}{A\arabic{table}}
\setcounter{table}{0}

\begin{table*}
    \centering
    \setlength\extrarowheight{5pt}
    \setlength\tabcolsep{2pt}
    \caption{Table 1 of FRB20250613 burst properties. MJD is measured at the centroid of the on-burst region. The `type' column indicates which bursts a single component (S) or multiple components (M). The spectral Luminosity for the MeerKAT bursts was measured by integrating SEFD frequency function and assuming the beamforming efficency, digital loss as well as the $a_{ib}$ and $a_{cb}$ coefficients were $\sim$1.0 \citep{geyer2021thousand, jankowski2023sample, tian2025meerkat} and the number of polarisation N$_{b}$ = 2. The spectral luminosity of multi component bursts was measured by summing together individual components. }
    \label{tab:burst_tab1}
    \begin{tabular}{cccccccccc}
    \hline
    Burst ID & MJD & Type & Width (ms) & Fluence \% & Spec-lumin (10$^{33}$ ergs/s/Hz) & Search DM (pc cm$^{-3}$) & SMDM (pc cm$^{-3}$) & $\nu_{c}$ (MHz) & $\Delta \nu$ (MHz) \\ \hline
        M1 & 60849.2669579520 & S & 25 &  & 0.011$\pm$0.002 & 173.4 & 173.4 & 887.95 & 64.1 \\ 
        M2 & 60850.2723215385 & S & 4.173 & 95 & 0.019$\pm$0.003 & 177.1 & 176.7$^{3.6}_{3.3}$ & 1297.69 & 363.63 \\ 
        M3 & 60850.2749973220 & S & 7.848 & 90 & 0.011$\pm$0.002 & 176 & 176.0 & 1130.08 & 329.36 \\ 
        M4 & 60850.2800178817 & S & 5.627 & 95 & 0.052$\pm$0.008 & 177.1 & 179.2$^{5.2}_{5.3}$ & 1176.9 & 310.9 \\ 
        M5 & 60850.2931271104 & S & 8.23 & 95 & 0.040$\pm$0.006 & 175.3 & 177.6$^{4.5}_{1.5}$ & 1123.81 & 280.04 \\ 
        M6A & 60850.3061295820 & M & 5.13 & 97.5 & 0.092$\pm$0.010 & 175 & 174.7$^{0.3}_{0.3}$ & 1494.97 & 412.12 \\ 
        M6B & 60850.3061295820 & M & 3.101 & 97.5 &  & 175 & 174.7$^{0.3}_{0.3}$ & 1321.93 & 485.68 \\ 
        M7 & 60850.3385142861 & S & 4.173 & 80 & 0.043$\pm$0.007 & 176.1 & 176.9$^{8.0}_{5.0}$ & 1300.2 & 216.51 \\ 
        M8 & 60850.3559177178 & M & 2.641 & 95 & 0.123$\pm$0.019 & 175.3 & 175.5$^{1.6}_{2.2}$ & 1088.7 & 283.7 \\ 
        M10 & 60850.3906822521 & S & 14.47 & 90 & 0.036$\pm$0.006 & 177.5 & 180.4$^{10.0}_{10.0}$ & 1335.72 & 85.27 \\ 
        M9 & 60850.3906227268 & S & 9.953 & 95 & 0.033$\pm$0.005 & 177.5 & 180.7$^{3.9}_{4.2}$ & 1144.29 & 456.42 \\ 
        M11 & 60850.4066484197 & S & 5.283 & 95 & 0.057$\pm$0.009 & 176.1 & 177.2$^{3.8}_{6.0}$ & 1438.54 & 300.94 \\ 
        M12 & 60850.4264681160 & M & 7.55 & 95 & 0.035$\pm$0.006 & 177.1 & 178.4$^{5.7}_{13.4}$ & 1202.81 & 414.63 \\ 
        M13 & 60850.4402878774 & S & 14.968 &  & 0.010$\pm$0.002 & 177.1 & 177.0$^{9.0}_{3.0}$ & 1286.41 & 366.14 \\ 
        M14 & 60850.4534522619 & S & 10.489 &  & 0.042$\pm$0.007 & 172.9 & 176.6$^{10.0}_{10.0}$ & 1367.91 & 149.63 \\ 
        M15A & 60850.4592417595 & S & 8.23 & 80 & 0.028$\pm$0.004 & 175 & 175.1$^{2.2}_{2.0}$ & 1031.02 & 134.59 \\ 
        M15B & 60850.4592417595 & S & 9.034 & 90 &  & 175 & 179.9$^{4.0}_{8.0}$ & 929.46 & 132.08 \\ 
        M16 & 60850.4912642470 & S & 2.3 & 85 & 0.022$\pm$0.004 & 177.1 & 177.1 & 993.83 & 215.67 \\ 
        M17 & 60850.4531752737 & S & 3.598 &  & 0.045$\pm$0.007 & 178.5 & 179.4$^{7.0}_{5.6}$ & 1311.06 & 221.52 \\ 
        M18 & 60850.4196334931 & S & 10.3 &  & 0.025$\pm$0.004 & 178.2 & 178.2 & 957.5 & 95 \\ 
        M19 & 60850.2646356144 & S & 9.32 &  & 0.009$\pm$0.002 & 178.2 & 178.2 & 1067.81 & 213.16 \\ 
        M20 & 60850.2647261807 & S & 11.7 &  & 0.005$\pm$0.001 & 177.1 & 177.1 & 1012.22 & 312.64 \\ 
        M21 & 60849.2792568586 & S & 10.2 &  & 0.014$\pm$0.003 & 179.3 & 179.3 & 1275.54 & 40.96 \\ 
        M22 & 60850.3243758926 & S & 14.6 &  & 0.002$\pm$0.001 & 172.2 & 172.2 & 1125.1 & 349 \\ 
        M23 & 60850.3258641507 & S & 4.89 &  & 0.013$\pm$0.002 & 175.3 & 175.3 & 1222.04 & 78.6 \\ 
        M24 & 60850.3436056676 & M & 19.98 & 97.5 & 0.129$\pm$0.020 & 175 & 174.7$^{0.1}_{0.1}$ & 952.86 & 193.94 \\ 
        A1A & 60839.9318351840 & S & 3.155 & 95 & 0.953$\pm$0.191 & 174.6 & 174.9$^{0.7}_{0.7}$ & 1005.5 & 164 \\ 
        A1B & 60839.9318351840 & S & 0.27 & 95 &  & 174.6 & 174.6$^{0.1}_{0.1}$ & 887.7 & 181.2 \\ 
        A2 & 60974.7048986863 & M & 0.604 & 99 & 0.628$\pm$0.126 & 174.4 & 174.4 & 759 & 127 \\ 
        P1 & 60905.7742059100 & S & 0.768 & 95 & 1.478$\pm$0.055 & 175 & 175.0$^{0.9}_{0.4}$ & 1009 & 69 \\ 
        P2 & 60905.7200659000 & S & 3.328 & 97.5 & 1.588$\pm$0.026 & 175 & 174.7$^{0.5}_{0.4}$ & 1055 & 174 \\ 
        P3 & 60970.6023090000 & S & 1.024 &  & 1.368$\pm$0.059 & 175 & 174.4$^{0.2}_{0.4}$ & 796.75 & 125.5 \\ 
        P4 & 60973.5473258200 & S & 0.512 &  & 3.922$\pm$0.094 & 175 & 175.0 & 1002.75 & 23.5 \\ 
        P5 & 60974.6564376700 & M & 17.152 &  & 0.755$\pm$0.010 & 175 & 174.5$^{1.2}_{0.7}$ & 827 & 246 \\ 
        P6 & 60974.6584522900 & S & 4.608 &  & 0.553$\pm$0.021 & 175 & 175.0 & 782 & 156 \\ 
        P7 & 60974.5988630800 & M & 7.168 &  & 0.401$\pm$0.025 & 175 & 174.4$^{1.6}_{1.1}$ & 776.5 & 87 \\ 
        P8 & 60987.4858621200 & M & 16.128 &  & 0.519$\pm$0.015 & 175 & 174.2$^{0.5}_{1.3}$ & 833.25 & 202 \\ 
        P9 & 60987.4638832800 & S & <=1.024 &  & 1.493$\pm$0.054 & 175 & 174.2$^{2.4}_{1.3}$ & 795 & 126 \\ 
        P10 & 60987.4806313700 & M & 8.96 &  & 0.565$\pm$0.025 & 175 & 174.4$^{5.6}_{4.4}$ & 775 & 82 \\ 
    \end{tabular}
\end{table*}

\begin{table*}
    \centering
    \setlength\extrarowheight{5pt}
    \caption{Table 2 of FRB20250613 burst properties. The second column shows the on-burst integrated S/N. $\tau$ measurements denoted by "*" were fitted using a sub-band of the burst. The central frequency and bandwidth of the sub-band is reported in columns 4 and 5. $\tau$ is measured at the central frequency of the burst. The `Channel Zapping' column represents bursts that underwent channel filtering for polarisation measurements assuming a $\sigma$ = 3.0 threshold. "*" denotes bursts that used a smaller $\sigma$ threshold for appropriate analysis.}
    \label{tab:burst_tab2}
    \begin{tabular}{cccccccccc}
    \hline
        Burst ID & S/N & $\tau$ (ms) & $\tau$ subband $\nu$ & $\tau$ subband $\delta \nu$ & RM (rad m$^{-2}$) & P/I & L/I & |V|/I & Channel Zapping \\ 
        \hline
        M1 & 13.77 & <= 1.1 &  &  &  &  &  &  & N \\ 
        M2 & 16.93 & 0.77$^{0.33}_{0.45}$ &  &  & -7132.0$\pm$4.5 & 1.13$\pm$0.09 & 1.03$\pm$0.08 & 0.21$\pm$0.07 & N \\ 
        M3 & 10.04 & 1.87$^{0.55}_{0.63}$ &  &  &  &  &  &  & N \\ 
        M4 & 31.44 & <= 0.436$^{0.38}_{0.29}$ &  &  & -7112.1$\pm$2.8 & 0.65$\pm$0.03 & 0.59$\pm$0.02 & 0.14$\pm$0.03 & Y \\ 
        M5 & 29.34 & <= 3.0$^{0.37}_{0.34}$ &  &  & -7169.1$\pm$6.1 & 0.48$\pm$0.03 & 0.33$\pm$0.03 & 0.31$\pm$0.03 & Y \\ 
        M6A & 93.74 & <= 0.06$^{0.06*}_{0.05}$ & 1448.5 & 38.5 & -7111.2$\pm$0.7 & 0.95$\pm$0.02 & 0.94$\pm$0.02 & 0.02$\pm$0.01 & Y \\ 
        M6B & 101.17 & <= 0.038$^{0.02*}_{0.02}$ & 1393 & 66 & -7119.0$\pm$0.3 & 0.84$\pm$0.01 & 0.83$\pm$0.01 & 0.02$\pm$0.01 & Y \\ 
        M7 & 25.89 & 1.16$^{0.19*}_{0.20}$ & 1349.2 & 34.4 & -7168.9$\pm$10.0 & 0.67$\pm$0.04 & 0.61$\pm$0.04 & 0.12$\pm$0.05 & Y \\ 
        M8 & 66.68 & 0.4$^{0.08*}_{0.09}$ & 1009.7 & 98.64 & -7169.2$\pm$6.1 & 0.25$\pm$0.02 & 0.20$\pm$0.02 & 0.06$\pm$0.02 & Y \\ 
        M10 & 32.34 & 1.21$^{0.29}_{0.40}$ &  &  & -7211.9$\pm$10.0 & 0.95$\pm$0.05 & 0.90$\pm$0.04 & 0.13$\pm$0.05 & Y \\ 
        M9 & 22.92 & 1.77$^{0.49*}_{0.56}$ & 1114.5 & 61 & -7105.2$\pm$2.5 & 0.38$\pm$0.03 & 0.34$\pm$0.03 & 0.08$\pm$0.03 & Y \\ 
        M11 & 39.61 & 0.67$^{0.11}_{0.14}$ &  &  & -7257.5$\pm$2.0 & 0.84$\pm$0.02 & 0.82$\pm$0.02 & 0.09$\pm$0.03 & Y \\ 
        M12 & 35.89 & 1.34$^{0.29*}_{0.28}$ & 1360.56 & 99.12 & -7085.3$\pm$9.5 & 1.04$\pm$0.04 & 1.00$\pm$0.04 & 0.12$\pm$0.04 & Y \\ 
        M13 & 17.27 & <= 2.4$^{0.50}_{0.40}$ &  &  &  &  &  &  & Y* \\ 
        M14 & 40.20 & <= 2.4$^{0.31*}_{0.30}$ & 1316.5 & 47 &  &  &  &  & Y \\ 
        M15A & 20.80 & 0.87$^{0.30}_{0.40}$ &  &  & -7085.2$\pm$12.7 & 0.34$\pm$0.05 & 0.30$\pm$0.05 & 0.15$\pm$0.06 & Y \\ 
        M15B & 20.06 & <= 2.12$^{0.60}_{0.80}$ &  &  &  &  &  &  & Y \\ 
        M16 & 8.35 & <= 0.75$^{0.55}_{0.48}$ &  &  &  &  &  &  & N \\ 
        M17 & 27.50 & 0.99$^{0.35*}_{0.49}$ & 1371.7 & 68.55 & -7188.8$\pm$11.0 & 0.18$\pm$0.03 & 0.14$\pm$0.03 & 0.09$\pm$0.04 & Y \\ 
        M18 & 14.75 &  &  &  &  &  &  &  & N \\ 
        M19 & 8.00 &  &  &  &  &  &  &  & N \\ 
        M20 & 4.68 &  &  &  &  &  &  &  & N \\ 
        M21 & 6.42 &  &  &  &  &  &  &  & N \\ 
        M22 & 2.97 &  &  &  &  &  &  &  & N \\ 
        M23 & 5.00 &  &  &  &  &  &  &  & N \\ 
        M24 & 168.30 &  &  &  & -7117.4$\pm$3.9 & 0.39$\pm$0.01 & 0.37$\pm$0.01 & 0.06$\pm$0.01 & Y \\ 
        A1A & 61.15 & 0.95$^{0.04}_{0.04}$ &  &  & -7215.8$\pm$5.4 & 0.19$\pm$0.01 & 0.17$\pm$0.01 & 0.05$\pm$0.02 & Y \\ 
        A1B & 46.73 & 0.05$^{0.01}_{0.01}$ &  &  & -7151.5$\pm$0.9 & 0.51$\pm$0.02 & 0.49$\pm$0.02 & 0.04$\pm$0.02 & Y \\ 
        A2 & 101.69 & 0.06$^{0.01}_{0.01}$ &  &  & -7365.6$\pm$0.8 & 0.30$\pm$0.01 & 0.28$\pm$0.01 & 0.05$\pm$0.01 & Y \\ 
        P1 & 26.98 &  &  &  & -6698.7$\pm$11.5 & 0.17$\pm$0.03 & 0.14$\pm$0.03 & 0.06$\pm$0.03 & Y \\ 
        P2 & 61.38 &  &  &  & -6626.1$\pm$2.7 & 0.46$\pm$0.02 & 0.46$\pm$0.02 & 0.03$\pm$0.02 & Y \\ 
        P3 & 23.42 &  &  &  & -7629.0$\pm$4.4 & 0.34$\pm$0.05 & 0.33$\pm$0.05 & 0.05$\pm$0.05 & N \\ 
        P4 & 42.05 &  &  &  & -7399.5$\pm$38.5 & 0.74$\pm$0.04 & 0.73$\pm$0.04 & 0.04$\pm$0.05 & N \\ 
        P5 & 81.14 &  &  &  & -7408.8$\pm$2.4 & 0.13$\pm$0.02 & 0.12$\pm$0.02 & 0.03$\pm$0.02 & Y* \\ 
        P6 & 26.51 &  &  &  & -7305.6$\pm$5.1 & 0.22$\pm$0.04 & 0.20$\pm$0.04 & 0.04$\pm$0.05 & N \\ 
        P7 & 16.27 &  &  &  &  &  &  &  & N \\ 
        P8 & 36.52 &  &  &  & -6773.6$\pm$2.0 & 0.26$\pm$0.03 & 0.25$\pm$0.03 & 0.02$\pm$0.03 & Y \\ 
        P9 & 27.67 &  &  &  & -6984.4$\pm$7.7 & 0.19$\pm$0.04 & 0.17$\pm$0.04 & 0.03$\pm$0.04 & Y \\ 
        P10 & 22.94 &  &  &  & -6951.8$\pm$12.4 & 0.12$\pm$0.04 & 0.12$\pm$0.04 & 0.01$\pm$0.05 & N \\ 
    \end{tabular}
\end{table*}

\subsection{Appendix: B - Additional plots}
\label{sec: appendixB}
\renewcommand{\thefigure}{B\arabic{figure}}
\setcounter{figure}{0}

\begin{figure}
    \centering
    \includegraphics[width=\linewidth]{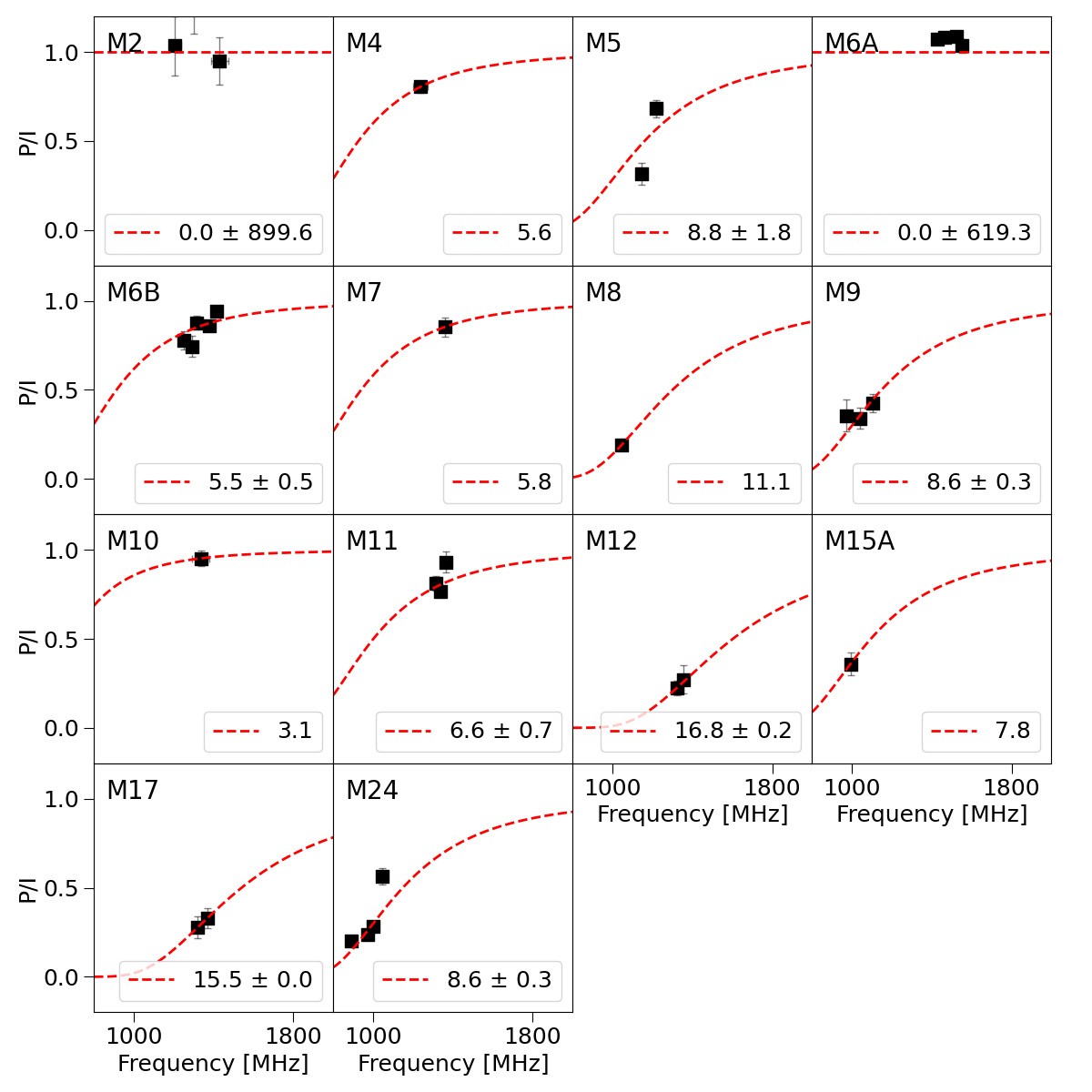}
    \caption{Spectral depolarisation fitting for MeerKAT bursts assuming $p$ = 1.0 in Eq \ref{eq: burnslaw}. }
    \label{fig:meerkat_p1.0_depol_mosaic}
\end{figure}

\begin{figure}
    \centering
    \includegraphics[width=\linewidth]{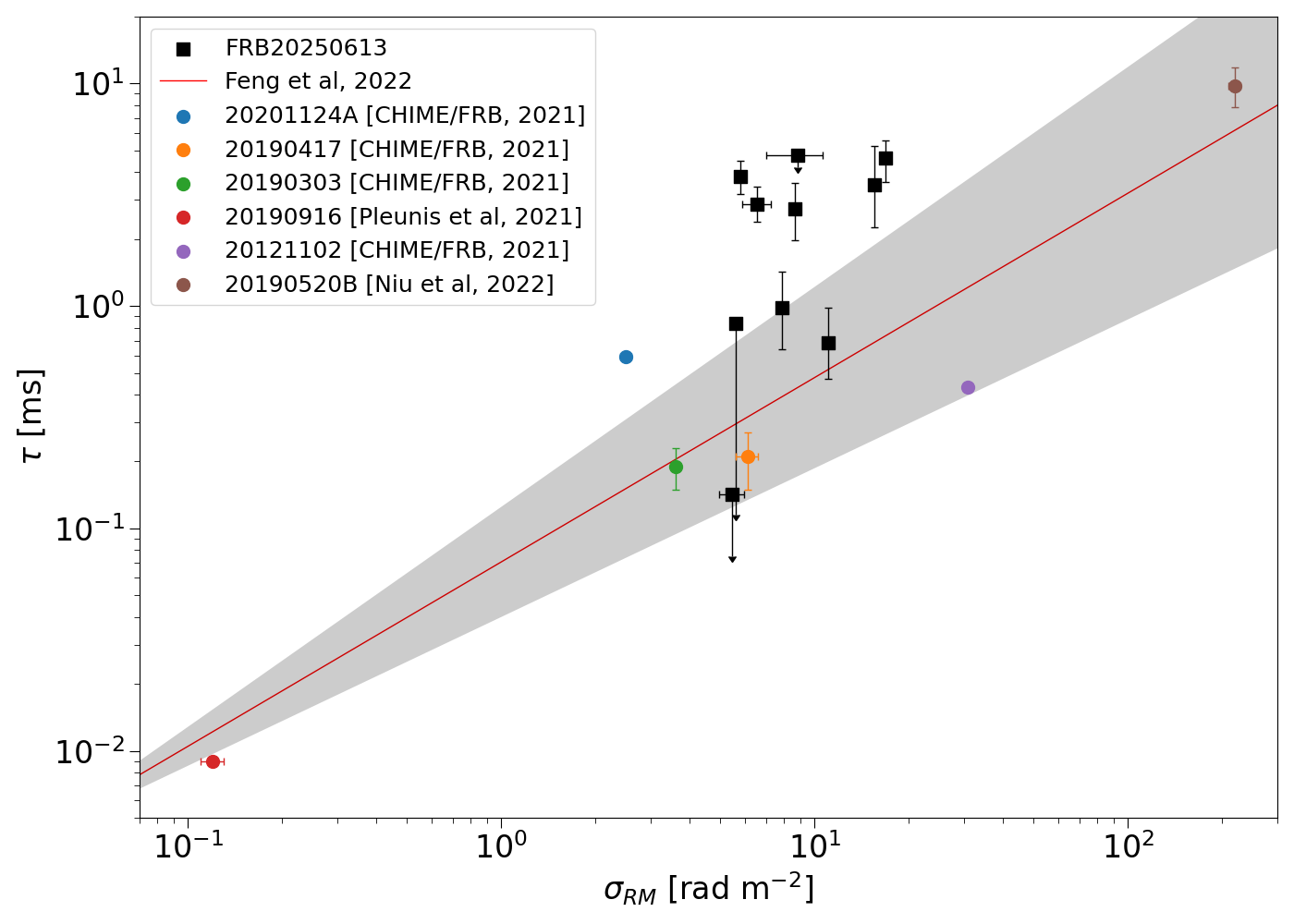}
    \caption{Observed relationship between $\mathrm{\sigma_{RM}}$ and $\tau$ for MeerKAT bursts compared against the empirical model from \citet{feng2022frequency}.}
    \label{fig:meerkat_tau_feng}
\end{figure}


\bsp	
\label{lastpage}
\end{document}